\documentclass[apj]{emulateapj}
\usepackage{natbib}

\newcommand{\rl}{$R_{\rm BLR}$--$L$}
\newcommand{\msigma}{$M_{\rm BH}-\sigma_{\star}$}
\newcommand{\mbh}{$M_{\rm BH}$}
\newcommand{\hst}{{\it HST}}
\newcommand{\sersic}{S\'{e}rsic}

\shorttitle{Black Hole Mass of MCG-06-30-15}
\shortauthors{Bentz, et al.}

\received{}
\accepted{}

\begin{document}

\title{A Reverberation-Based Black Hole Mass for MCG-06-30-15}

\author{ Misty~C.~Bentz\altaffilmark{1},
Edward~M.~Cackett\altaffilmark{2},
D.~Michael~Crenshaw\altaffilmark{1}, \linebreak
Keith~Horne\altaffilmark{3},
Rachel~Street\altaffilmark{4},
and
Benjamin~Ou-Yang\altaffilmark{1}}

\altaffiltext{1}{Department of Physics and Astronomy,
		 Georgia State University,
		 Atlanta, GA 30303, USA;
		 bentz@astro.gsu.edu}

\altaffiltext{2}{Department of Physics and Astronomy,
                 Wayne State University,
                 666 W.\ Hancock St.,
                 Detroit, MI 48201, USA}

\altaffiltext{3}{SUPA Physics and Astronomy, 
                 University of St. Andrews,
                 Fife, KY16 9SS Scotland, UK}

\altaffiltext{4}{Las Cumbres Observatory Global Telescope Network,
                 6740B Cortona Drive,
                 Goleta, CA 93117, USA}

\begin{abstract}
We present the results of a reverberation campaign targeting
MGC-06-30-15.  Spectrophotometric monitoring and broad-band
photometric monitoring over the course of 4 months in the spring of
2012 allowed a determination of a time delay in the broad H$\beta$
emission line of $\tau=5.3\pm1.8$ days in the rest frame of the
AGN. Combined with the width of the variable portion of the emission
line, we determine a black hole mass of $M_{\rm BH} = (1.6 \pm 0.4)
\times 10^6$\,M$_{\odot}$.  Both the H$\beta$ time delay and the black
hole mass are in good agreement with expectations from the \rl\ and
\msigma\ relationships for other reverberation-mapped AGNs.  The
H$\beta$ time delay is also in good agreement with the relationship
between H$\beta$ and broad-band near-IR delays, in which the effective 
BLR size is $\sim 4-5$ times smaller than the inner edge of the dust
torus.  Additionally, the reverberation-based mass is in good
agreement with estimates from the X-ray power spectral density break
scaling relationship, and with constraints based on stellar kinematics
derived from integral field spectroscopy of the inner $\sim 0.5$\,kpc
of the galaxy.
\end{abstract}


\section{Introduction}

It has been a century since Edward \citet{fath13} first observed
strong emission lines originating in the nucleus of NGC\,1068 and
discovered the first active galactic nucleus (AGN).  Yet it is only in
the last 30 years that AGNs have become synonymous with supermassive
black holes (e.g., \citealt{rees84}) and that supermassive black holes
have become synonymous with galaxy nuclei (e.g.,
\citealt{magorrian98,ferrarese00,gebhardt00,ferrarese05}).  Multiple
independent lines of study focusing on a zoo of seemingly unrelated
characteristics across the entire spectral energy distribution are now
unified through our current understanding of the AGN phenomenon (e.g.,
\citealt{antonucci93,urry95}).  The mass and the spin of the black
hole, its only quantifiable characteristics, are two key parameters in our
understanding of not only AGN physics (e.g.,
\citealt{krawczynski13,netzer15}), but also galaxy evolution (e.g.,
\citealt{fabian12,kormendy13,heckman14,king15}).


Astrophysical black holes can be characterized by their mass and spin,
and being able to constrain both properties is rare.  MCG-06-30-15 is
one of only a handful of X-ray bright AGNs where its Fe K$\alpha$
emission may be studied in detail, allowing a measure of the black
hole spin. Tanaka et al. (1995) first detected a broad red wing of the
Fe K$\alpha$ emission line, as is expected due to the strong
gravitational redshift and relativistic Doppler effects from material
in the innermost accretion disk.  Relativistic reflection models fit
to the X-ray spectrum, including the Fe K$\alpha$ line, all indicate
that the black hole spin is high, with a dimensionless spin parameter
$a \gtrsim 0.9$ \citep{brenneman06, chiang11, marinucci14}.


\begin{figure*}
\epsscale{1.15}
\plottwo{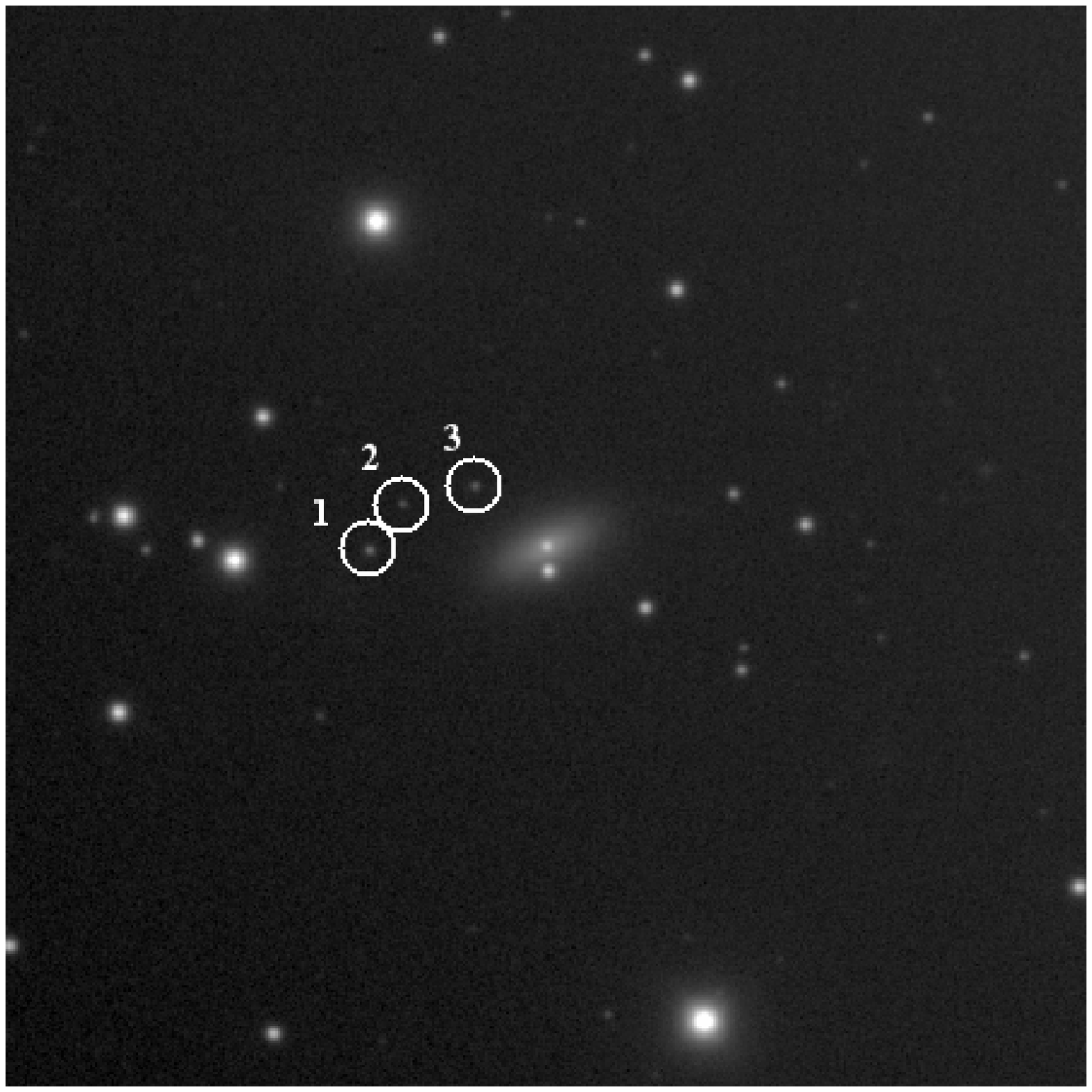}{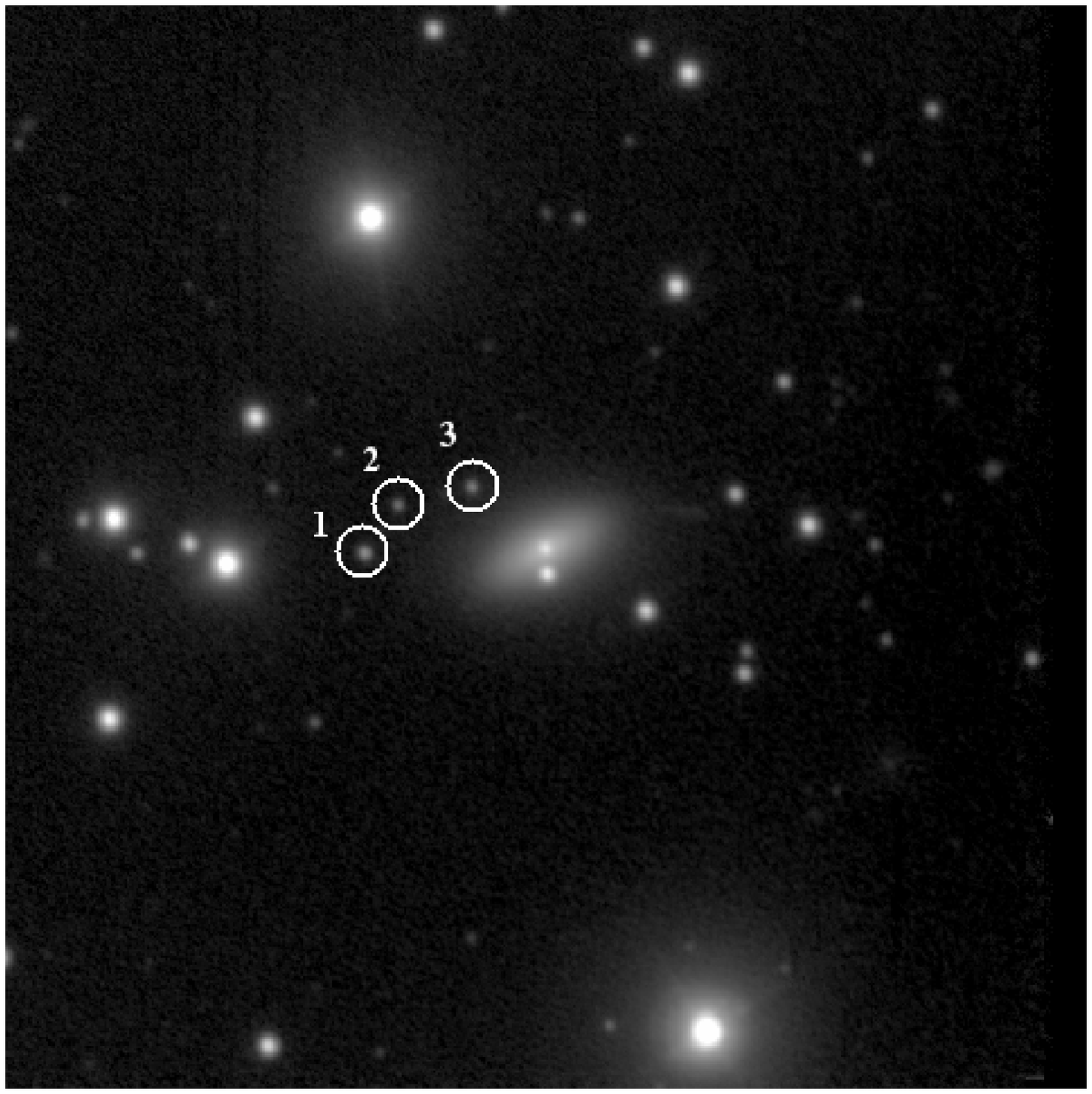}
\caption{Reference images for the $V-$band FTS dataset ({\it left})
  and the $V-$band LT dataset ({\it right}).  The region displayed is
  4\arcmin$\times$4\arcmin, with north up and east to the left.  The
  average exposure time for the FTS images is 20\,s at a typical
  airmass of 1.3, compared to 60\,s exposures for the LT images with
  typical airmasses of 2.3.  The marked field stars are common to the
  ground-based images and the high-resolution \hst\ image and were
  used to determine the absolute flux scale of the FTS and LT images.}
\label{fig:refims}
\end{figure*}

While the spin has been constrained for a decade now, the black hole
mass of MCG-06-30-15 is less well known.  Previous estimates of the
mass have relied upon scaling relationships such as the X-ray power
spectral density break ($M_{\rm BH} = 2.9^{+1.8}_{-1.6} \times
10^6$\,M$_{\odot}$; \citealt{mchardy05}) or the \msigma\ relationship
($M_{\rm BH} = 3-6 \times 10^6$\,M$_{\odot}$; \citealt{mchardy05}).
High spatial resolution integral field spectroscopy of the inner $\sim
0.5$\,kpc of the galaxy allowed \citet{raimundo13} to determine an
upper limit on the black hole mass of $<6 \times 10^7$\,M$_{\odot}$,
but the integration time was somewhat shallow and precluded a stronger
mass constraint.

Reverberation mapping \citep{blandford82,peterson93} is often employed
for determining the black hole masses of AGNs of interest.  Unlike
dynamical modeling, which is limited by spatial resolution and
therefore distance, reverberation mapping is applicable to all
broad-lined AGNs regardless of location.  The method makes use of the
spectral variability of AGNs and determines the time delay between
variations in the continuum emission (likely emitted from the
accretion disk) and the response to these variations in the broad
emission lines (emitted from the broad line region, BLR).  The time
delay is simply the responsivity-weighted average of the light travel
time from the accretion disk to all of the BLR ``clouds'', and is
generally interpreted as a measure of the average radius of the BLR
for a specific emission species.  In this case, the limiting
resolution is temporal rather than spatial, and regions on the order
of microarcseconds in size are routinely investigated (e.g.,
\citealt{peterson02,bentz09c,denney10,grier12b}).  The time delay
combined with a measure of the velocity of the gas provides a
constraint on the black hole mass through the virial theorem, modulo a
scaling factor that accounts for the detailed geometry and kinematics
of the line-emitting gas.

The requirements of dense temporal sampling and long monitoring
baselines have generally limited reverberation campaigns to 1.0-4.0\,m
class telescopes in the past, and these have generally been located in
the Northern Hemisphere.  At a declination of $\delta = -34.3\degr$,
MCG-06-30-15 has not been an ideal target for a reverberation
campaign.  Nevertheless, it was included in the set of AGNs monitored
from Lick Observatory as part of the LAMP 2008 program
\citep{bentz09c}, but no time delays were detected due to the low
level of variability of the source throughout the campaign combined
with the non-optimal conditions under which it was observed each night
(airmass $>3$).

We describe here the results of a reverberation-mapping campaign for
MCG-06-30-15 anchored by spectroscopy from the SMARTS 1.5-m telescope
at Cerro Tololo Interamerican Observatory (CTIO).  The variability of
the target was somewhat increased during the monitoring period,
compared to the 2008 campaign, and coupled with better data quality,
we are able to determine a time delay for the broad H$\beta$ emission
line and a constraint on the black hole mass.

Throughout this paper, we adopt a $\Lambda$CDM cosmology of $H_0 =
72$\,km\,s$^{-1}$\,Mpc$^{-1}$, $\Omega_{\rm M}= 0.3$,
$\Omega_{\Lambda} = 0.7$.

\section{Observations}

For the monitoring campaign presented here, observations were carried
out during Spring 2012 with spectroscopy obtained at CTIO
(latitude\,$=-30\degr$), and photometry obtained at Siding Spring
Observatory (latitude\,$=-31\degr$) and at the Observatorio del Roque
de los Muchachos at La Palma (latitude\,$=+28\degr$).  The details of
each are described below.

\subsection{Photometry}

For reverberation mapping campaigns, photometric monitoring can
provide a higher signal-to-noise ratio and better calibrated light
curve of continuum variations than measurements taken directly from
the spectra (e.g., \citealt{bentz08,bentz09c}).  In the cases of the
$B$ and $V$ bands, especially, the contribution of broad-line emission
to the bandpass is small compared to the continuum (e.g.,
\citealt{walsh09}).  We therefore carried out broad-band $B$ and $V$
photometric monitoring at two sites to better constrain the continuum
variations throughout our campaign: the 2-m Faulkes Telescope South
(FTS) at Siding Spring Observatory and the 2-m Liverpool Telescope
(LT) at the Observatorio del Roque de los Muchachos on La Palma in the
Canary Islands.

Monitoring at FTS began on 4 February and continued through 26 May
2012 with the Spectral camera (UT dates here and throughout).
Observations were obtained on 42 nights and generally consisted of
$2\times35$\,s exposures in $B$ and $2\times20$\,s exposures in $V$ at
an average airmass of 1.3.  The field-of-view for the images was
10\farcm5$\times$10\farcm5 with a pixel scale of 0.304'' in $2\times2$
binning mode.

Monitoring at LT utilized the RATCam and was carried out from 20
February through 29 May 2012.  Observations were obtained on 42 nights
at a typical airmass of 2.27.  Because of the higher expected airmass
for these observations based on the latitude of the observatory
relative to the declination of the target, longer exposure times of
$2\times60$\,s were utilized in $B$ and $V$.  The field of view for
RATCam is 4\farcm6$\times$4\farcm6, with a pixel scale of 0\farcs28 in
$2\times2$ binning mode.

Both imaging datasets were analyzed through image subtraction methods
in order to accurately constrain the nuclear variability of the
galaxy.  Images from a single observatory and a single filter were
registered to a common alignment using the {\tt Sexterp} routine
\citep{siverd12}.  We then employed the {\tt ISIS} image subtraction
package \citep{alard98,alard00} to build a reference image
(Figure~\ref{fig:refims}) from the subset of images taken under the
best conditions.  The reference frame was convolved with a
spatially-varying kernel to match each individual image in the
dataset.  Subtraction of the convolved reference from each image
produces a residual image in which the components that are constant in
flux have disappeared and only variable sources remain.  Aperture
photometry was then employed on the residual images to measure the
amount of variable flux for the AGN.  Analysis of these resultant
light curves demonstrated that the $V-$band light curves exhibit the
same features as the $B-$band light curves, but with less noise.  We
therefore focus our remaining analysis on the $V-$band light curves
from our photometric monitoring.

\begin{figure}
\epsscale{1.15}
\plotone{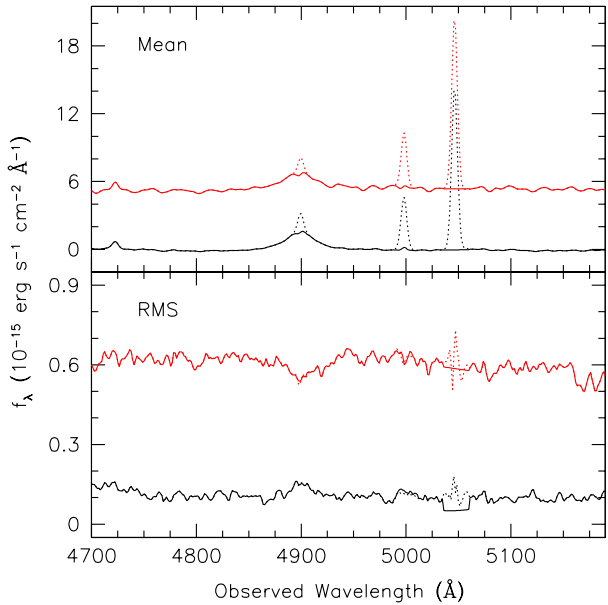}
\caption{{\it Top:} Mean spectrum of MCG-06-30-15 (red line) and the
  mean spectrum created after subtraction of the continuum from each
  individual spectrum (black line).  The dotted lines show the narrow
  component of H$\beta$ and the [\ion{O}{3}]
  $\lambda\lambda$\,4959,5007 doublet.  {\it Bottom:} RMS of the
  original calibrated spectra (red line) and rms of the
  continuum-subtracted spectra (black line).  The variable portion of
  the H$\beta$ emission line is apparent once the strongly variable
  continuum has been removed.}
\label{fig:meanspec}
\end{figure}

\subsection{Spectroscopy}

\begin{figure}
\epsscale{1.15}
\plotone{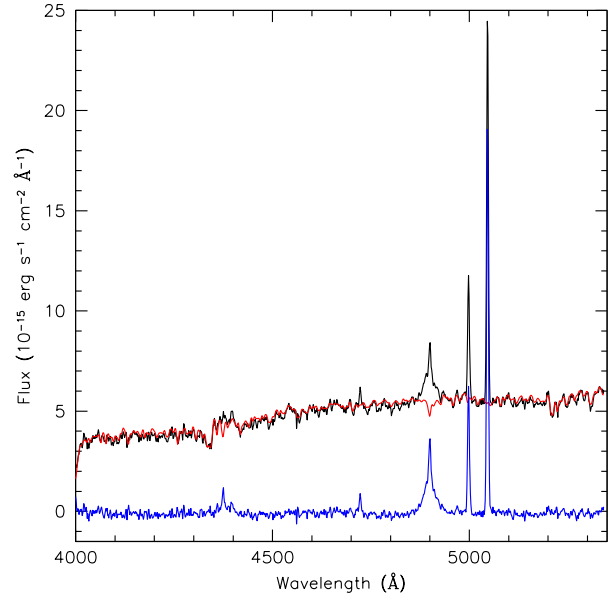}
\caption{The black line displays a spectrum of MCG-06-30-15 taken on
  the night of 1 March 2012, with the best-fit continuum model
  overplotted in red.  The continuum model is comprised of a powerlaw
  component and a model host-galaxy component.  The
  continuum-subtracted spectrum is plotted in blue.}
\label{fig:galsub}
\end{figure}

Spectroscopic monitoring was carried out with RCSpec on the SMARTS
1.5-m telescope at CTIO.  Observations were scheduled to be carried
out in queue-observing mode every other night during the period 1
March $-$ 31 May 2012.  The spectrograph was equipped with the
600\,l/mm blue grating (known as grating 26), giving a wavelength
coverage of $3685-5400$\,\AA\ and a nominal resolution of
1.5\,\AA\,pix$^{-1}$ in the dispersion direction.  Spectra were
obtained through a 4\arcsec\ slit at a fixed position angle of
90\degr\ (i.e., oriented east-west).  The RCSpec detector, a Loral 1K
CCD, provides a spatial resolution of 1\farcs3\,pix$^{-1}$.

Over the course of the campaign, spectra were obtained on 36
nights.  Each visit consisted of two spectra with exposure times of
900\,s that were obtained at an average airmass of 1.08.  A
spectrophotometric flux standard, LTT\,4364, was also observed during
each visit to assist with flux calibrations.  Standard reductions were carried
out with IRAF\footnote{IRAF is distributed by the National Optical
  Astronomy Observatory, which is operated by the Association of
  Universities for Research in Astronomy (AURA) under cooperative
  agreement with the National Science Foundation.} and an extraction
width of 8\,pixels (10\farcs4) was adopted.

\begin{figure*}
\epsscale{1.1}
\plotone{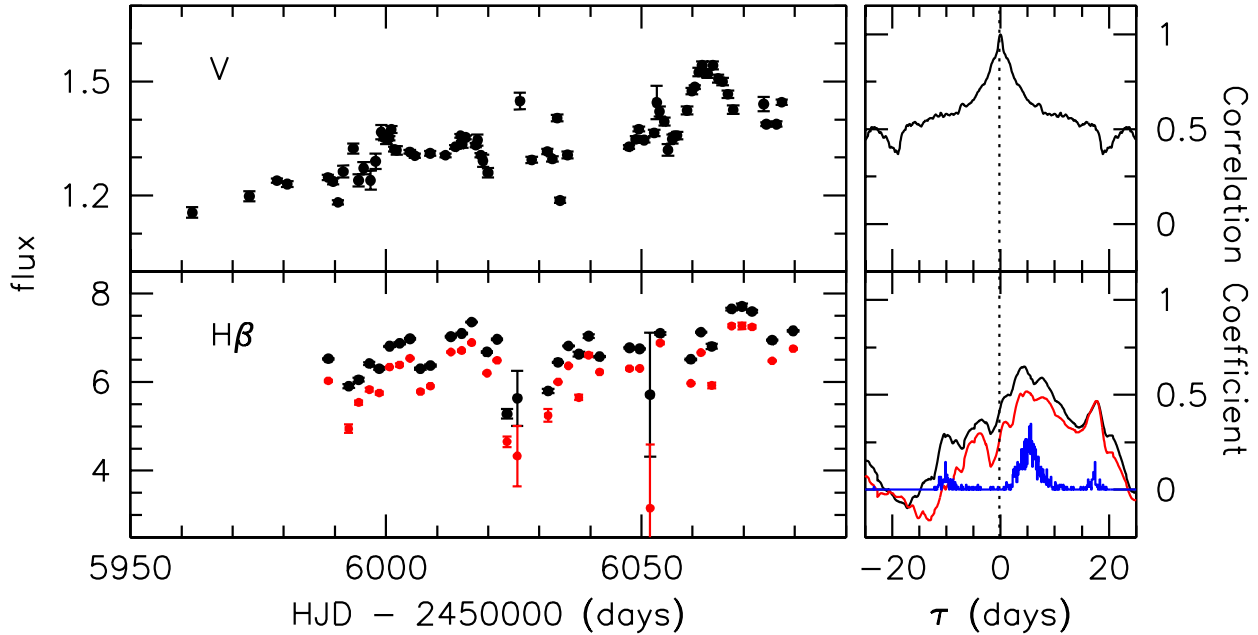}
\caption{Continuum light curve ({\it top left}) and its
  autocorrelation function ({\it top right}), and the H$\beta$
  emission-line ({\it bottom left}) light curve and its
  cross-correlation function relative to the continuum ({\it bottom
    right}). Flux units in the light curves are as listed in
  Table~\ref{tab:lcdata}. In the bottom left panel, the black (red)
  points show the H$\beta$ emission-line flux measurements based on
  the continuum-subtracted spectra (original calibrated spectra).  The
  resultant cross-correlation function relative to the continuum light
  curve for each emission-line light curve is displayed in the same
  color in the bottom right panel.  The H$\beta$ light curves and
  cross-correlation functions are very similar, demonstrating that our
  continuum subtraction method does not bias our measurements.  In
  both cases, the emission-line cross-correlation function is an
  obviously shifted and smoothed version of the continuum
  auto-correlation function.  The blue histogram in the bottom right
  panel displays the (arbitrarily scaled) cross-correlation centroid
  distributions based on the continuum light curve and the H$\beta$
  light curve measured from the continuum-subtracted spectra, which
  gives an H$\beta$ time delay of $\tau = 5.4 \pm 1.8$\,days.}
\label{fig:lc}
\end{figure*}

The initial flux calibration provided by the standard star is
generally a good correction for the shape of the spectra, providing a
useful way to remove the effects of the atmospheric transmission as
well as the optics of the telescope and instrument.  However,
reverberation campaigns require high temporal sampling and therefore
acquire spectra on all nights when the telescope may be safely used,
often times under non-photometric conditions.  We therefore require a
method for carefully calibrating the overall flux level of each
spectrum.  This is generally accomplished by using the narrow emission
lines as ``internal'' flux calibration sources, as the narrow lines do
not vary on the timescales of a reverberation campaign.  Specifically,
we employ the \citet{vangroningen92} spectral scaling method with the
[\ion{O}{3}] $\lambda\lambda$\,4959,5007 emission lines as our
internal calibration sources.  The method minimizes the differences in
a selected wavelength range between each individual spectrum and a
reference spectrum created from a subset of the best data.  It is
therefore able to correct for slight differences in wavelength
calibration, slight resolution differences (caused by variable seeing
and the employment of a wide spectroscopic slit), as well as flux
calibration differences. \citet{peterson98a} have shown that this
method is able to provide relative spectrophotometry that is accurate
to $\sim 2$\%.  To ensure the accuracy of our absolute spectrophotometry,
we compared the integrated [\ion{O}{3}] $\lambda 5007$ flux to
published values determined from high quality spectra observed under
good conditions.  We adopted a value of $f($[\ion{O}{3}]$) = 1.0 \times
10^{-13}$\,erg\,s$^{-1}$\,cm$^{-2}$, in good agreement with
\citet{morris88}, \citet{winkler92}, and \citet{reynolds97}.

The red lines in Figure~\ref{fig:meanspec} show the mean of all the
calibrated spectra throughout the campaign ({\it top}) and the root
mean square of the spectra ({\it bottom}), which highlights variable
spectral components.  It is immediately obvious that the variable
(rms) spectrum is swamped by some combination of host-galaxy light,
possibly from mis-centering of the slit and poor seeing conditions, as
well as scattered light from a nearby Milky Way star along the line of
sight (5\farcs4 south of the nucleus and superimposed on the galaxy
disk).  We therefore investigated a method for carefully subtracting
the continuum of each spectrum, both the AGN powerlaw and the
starlight, through spectral modeling.

We employed the publicly-available {\tt UlySS} package
\citep{koleva09}, which creates a linear combination of non-linear
model components convolved with a parametric line-of-sight velocity to
match an observed spectrum.  Our method started with modeling the very
high signal-to-noise mean spectrum.  We included a powerlaw component
for the AGN continuum emission, multiple Gaussians for the emission
lines (three components were necessary to match the H$\beta$ profile),
and a host-galaxy component parametrized by the Vazdekis models
derived from the MILES library of empirical stellar spectra
\citep{vazdekis10}.  We make no attempt to interpret the best-fit
parameters of our model, as our goal was simply to separate the line
emission from the continuum components as cleanly as possible.  We
then held the number of model components and the age and metallicity
of the best-fit Vazdekis model fixed, but allowed all other parameters
to vary as we looped through all of the individual spectra of
MCG-06-30-15.  In this way, we allow for variation of the powerlaw
index, the relative contribution of powerlaw versus host-galaxy
starlight, and emission line flux variability.  Furthermore, we
modeled the AGN spectra that had an initial flux calibration from a
spectrophotometric standard star but had not yet been scaled with the
\citet{vangroningen92} code in order to get the best match between the
models and the ``untouched'' observed spectra.  The best-fit powerlaw
and host-galaxy models were subtracted from each spectrum, providing a
set of continuum subtracted, pure emission-line spectra.  In
Figure~\ref{fig:galsub}, we show a typical example of a single
spectrum, the best-fit continuum model (powerlaw + starlight), and the
resultant continuum-subtracted spectrum.

After modeling and subtraction of the continuum, all the spectra were
then scaled with the \citet{vangroningen92} method in the same way as
previously described.  The mean and rms of the scaled,
continuum-subtracted spectra are displayed by the black lines in
Figure~\ref{fig:meanspec}.  The H$\beta$ emission line, though weak,
is apparent in the continuum-subtracted rms spectrum.

\begin{deluxetable*}{cccc}
\tablecolumns{4}
\tablewidth{0pt}
\tablecaption{$V-$band and H$\beta$ Light Curves}
\tablehead{
\colhead{HJD} &
\colhead{$f_{\lambda}$(V)} &
\colhead{HJD} &
\colhead{$f$(H$\beta$)}\\
\colhead{(-2450000)} &
\colhead{($10^{-15}$ \,ergs\,s$^{-1}$\,cm$^{-2}$\,\AA$^{-1}$)} &
\colhead{(-2450000)} &
\colhead{($10^{-15}$ \,ergs\,s$^{-1}$\,cm$^{-2}$)} 
}
\startdata
5962.0886  &   $1.155 \pm\ 0.014$  &  5988.7106   & $65.27 \pm\	0.19$  \\     
5973.2684  &   $1.198 \pm\ 0.013$  &  5992.7187   & $59.09 \pm\	0.43$  \\      
5978.6867  &   $1.239 \pm\ 0.006$  &  5994.6647   & $60.54 \pm\	0.44$  \\      
5980.6836  &   $1.230 \pm\ 0.008$  &  5996.7522   & $64.18 \pm\	0.34$  \\      
5988.6514  &   $1.248 \pm\ 0.007$  &  5998.6879   & $63.05 \pm\	0.31$  \\      
5989.6408  &   $1.237 \pm\ 0.008$  &  6000.7037   & $68.11 \pm\	0.12$  \\      
5990.6260  &   $1.182 \pm\ 0.006$  &  6002.6755   & $68.78 \pm\	0.28$  \\      
5991.6260  &   $1.263 \pm\ 0.016$  &  6004.6979   & $69.77 \pm\	0.16$  \\      
5993.6192  &   $1.324 \pm\ 0.013$  &  6006.7308   & $63.05 \pm\	0.21$  \\      
5994.6396  &   $1.240 \pm\ 0.016$  &  6008.6991   & $63.75 \pm\	0.22$  \\      
5995.6130  &   $1.272 \pm\ 0.016$  &  6012.7055   & $70.27 \pm\	0.17$  \\      
5996.9981  &   $1.240 \pm\ 0.025$  &  6014.8060   & $70.97 \pm\	0.22$  \\      
5997.9778  &   $1.290 \pm\ 0.020$  &  6016.7763   & $73.57 \pm\	0.13$  \\      
5999.0720  &   $1.367 \pm\ 0.019$  &  6019.7684   & $66.81 \pm\	0.17$  \\      
5999.6167  &   $1.353 \pm\ 0.007$  &  6021.7506   & $69.67 \pm\	0.15$  \\      
6000.0813  &   $1.353 \pm\ 0.017$  &  6023.6823   & $52.83 \pm\	1.06$  \\      
6000.6030  &   $1.351 \pm\ 0.006$  &  6025.6963   & $56.35 \pm\	6.20$  \\      
6001.0505  &   $1.374 \pm\ 0.009$  &  6031.7218   & $58.02 \pm\	0.48$  \\      
6001.6535  &   $1.322 \pm\ 0.006$  &  6033.6779   & $64.49 \pm\	0.13$  \\      
6002.0935  &   $1.319 \pm\ 0.011$  &  6035.6990   & $68.14 \pm\	0.11$  \\      
6004.6577  &   $1.315 \pm\ 0.005$  &  6037.7316   & $66.33 \pm\	0.53$  \\      
6005.6684  &   $1.305 \pm\ 0.006$  &  6039.6459   & $70.39 \pm\	0.39$  \\      
6008.6396  &   $1.311 \pm\ 0.006$  &  6041.8089   & $65.73 \pm\	0.14$  \\      
6011.6284  &   $1.306 \pm\ 0.005$  &  6047.6504   & $67.73 \pm\	0.17$  \\      
6013.5708  &   $1.328 \pm\ 0.005$  &  6049.6533   & $67.52 \pm\	0.14$  \\      
6014.5952  &   $1.357 \pm\ 0.005$  &  6051.6786   & $57.17 \pm\	14.03$ \\       
6014.9191  &   $1.335 \pm\ 0.009$  &  6053.6594   & $70.99 \pm\	0.29$  \\      
6015.5847  &   $1.353 \pm\ 0.005$  &  6059.6553   & $65.15 \pm\	0.17$  \\      
6017.5606  &   $1.333 \pm\ 0.006$  &  6061.6300   & $71.26 \pm\	0.14$  \\      
6017.9114  &   $1.346 \pm\ 0.015$  &  6063.7588   & $68.08 \pm\	0.56$  \\      
6018.5535  &   $1.306 \pm\ 0.006$  &  6067.6406   & $76.50 \pm\	0.29$  \\      
6018.9996  &   $1.291 \pm\ 0.016$  &  6069.6631   & $77.08 \pm\	0.68$  \\      
6019.9486  &   $1.260 \pm\ 0.012$  &  6071.6367   & $75.97 \pm\	0.28$  \\      
6026.2009  &   $1.449 \pm\ 0.022$  &  6075.5645   & $69.45 \pm\	0.14$  \\      
6028.5347  &   $1.294 \pm\ 0.008$  &  6079.6835   & $71.57 \pm\	0.17$  \\      
6031.5757  &   $1.316 \pm\ 0.007$  &    &   \\
6032.5654  &   $1.296 \pm\ 0.007$  &    &   \\
6033.5161  &   $1.404 \pm\ 0.009$  &    &   \\
6034.0558  &   $1.187 \pm\ 0.007$  &    &   \\
6035.5098  &   $1.307 \pm\ 0.009$  &    &   \\
6047.5410  &   $1.328 \pm\ 0.005$  &    &   \\
6048.8850  &   $1.348 \pm\ 0.009$  &    &   \\
6049.4712  &   $1.375 \pm\ 0.006$  &    &   \\
6050.5168  &   $1.346 \pm\ 0.006$  &    &   \\
6052.4839  &   $1.365 \pm\ 0.008$  &    &   \\
6052.9778  &   $1.445 \pm\ 0.044$  &    &   \\
6053.5425  &   $1.421 \pm\ 0.013$  &    &   \\
6054.4863  &   $1.395 \pm\ 0.010$  &    &   \\
6055.1726  &   $1.320 \pm\ 0.015$  &    &   \\
6056.0708  &   $1.349 \pm\ 0.012$  &    &   \\
6056.4541  &   $1.357 \pm\ 0.006$  &    &   \\
6056.9783  &   $1.358 \pm\ 0.010$  &    &   \\
6058.9017  &   $1.424 \pm\ 0.010$  &    &   \\
6059.8591  &   $1.475 \pm\ 0.008$  &    &   \\
6060.4424  &   $1.486 \pm\ 0.005$  &    &   \\
6061.1837  &   $1.525 \pm\ 0.011$  &    &   \\
6061.8630  &   $1.543 \pm\ 0.010$  &    &   \\
6062.9246  &   $1.521 \pm\ 0.011$  &    &   \\
6063.9783  &   $1.543 \pm\ 0.010$  &    &   \\
6065.0056  &   $1.508 \pm\ 0.011$  &    &   \\
6065.8806  &   $1.500 \pm\ 0.010$  &    &   \\
6066.8920  &   $1.467 \pm\ 0.010$  &    &   \\
6067.9285  &   $1.426 \pm\ 0.011$  &    &   \\
6073.9258  &   $1.441 \pm\ 0.019$  &    &   \\
6074.4324  &   $1.388 \pm\ 0.005$  &    &   \\
6076.4253  &   $1.388 \pm\ 0.007$  &    &   \\
6077.4556  &   $1.446 \pm\ 0.007$  &    &   \\

\label{tab:lcdata}
\enddata 
\end{deluxetable*}

\section{Light Curve Analysis}

Emission-line light curves were determined from both the original,
scaled spectra as well as the continuum-subtracted, scaled spectra, in
order to verify that our continuum subtraction method did not
introduce artificial variability.  In both cases, a local linear
continuum was fit underneath the emission line, and the flux above the
continuum was integrated.  We included this local continuum fit even
for the continuum-subtracted spectra to ensure that any small
mismatches between the model continuum and that of the spectrum were
accounted for and removed from the emission-line
measurements. Multiple measurements from a single night were then
averaged together to decrease the noise in the resultant light curves,
which are displayed in Figure~\ref{fig:lc}.  The H$\beta$ light curve
derived from the scaled, continuum-subtracted spectra ({\it black
  points}) matches extremely well with the H$\beta$ light curve
derived from the scaled-only spectra ({\it red points}).  The two
light curves are virtually identical, with the most obvious difference
being a slight offset in which the continuum-subtracted spectra have
an elevated H$\beta$ flux (due to correction of the intrinsic H$\beta$
absorption from the starlight).  A linear fit to the fluxes determined
from each method shows that the difference between the two lightcurves
is almost entirely a simple offset, with very minimal flux dependence
(close to a slope of 1).

The differential light curves derived from image subtraction analysis
of the $V-$band photometry were converted to absolute flux units in
the following way.  First, the reference image for each set of
observations was modeled with the two-dimensional surface brightness
fitting program {\tt GALFIT}.  The shape parameters of the galaxy
bulge and disk were matched to those derived from the analysis of a
high-resolution medium$-V$ {\it Hubble Space Telescope (HST)} image
(see Section\,6.1).  Field stars common to both the {\it HST} image
and the ground-based images were also modeled (circled in
Figure~\ref{fig:refims}), and the field star magnitudes derived from
the {\it HST} image were used to set the absolute flux calibration of
the ground-based images.  The brightness of the AGN point spread
function in each ground-based reference image was then added back to
the differential flux derived from the image subtraction analysis for
that set of photometry.  While the overall flux scale of the light
curve is not important, we found a slight offset of 0.2\,mag between
the FTS and the LT calibrated photometry, so we adjusted the LT
photometry to match that of the FTS, since the FTS observations were
generally obtained under better conditions.  The calibrated
photometric light curves were then combined together and measurements
coincident within 0.5\,days were averaged together.  The final
$V-$band light curve is displayed in the top left panel of
Figure\,\ref{fig:lc}.

Table~\ref{tab:lcstats} gives the variability statistics for the final
$V-$band and H$\beta$ emission-line light curves displayed in
Figure~\ref{fig:lc}.  Column (1) lists the spectral feature and column
(2) gives the number of measurements in the light curve.  Columns (3)
and (4) list the average and median time separation between
measurements, respectively.  Column (5) gives the mean flux and
standard deviation of the light curve, and column (6) lists the mean
fractional error (based on the comparison of observations that are
closely spaced in time).  Column (7) lists the fractional rms variability amplitude,
computed as:
\begin{equation}
F_{\rm var} = \frac{\sqrt{\sigma^2 - \delta^2}}{\langle F \rangle}
\end{equation}
\noindent where $\sigma^2$ is the variance of the fluxes, $\delta^2$
is their mean-square uncertainty, and $\langle F \rangle$ is the mean
flux \citep{rodriguez97}. The uncertainty on $F_{\rm var}$ is quantified as
\begin{equation}
  \sigma_{F_{\rm var}} = \frac{1}{F_{\rm var}} \sqrt{\frac{1}{2N}} \frac{\sigma^2}{\langle F \rangle^2}
\end{equation}
\noindent \citep{edelson02}.  And column (8) is the ratio of the
maximum to the minimum flux in the light curve.  At first glance, the
$F_{\rm var}$ values for the H$\beta$ light curves from the
continuum-subtracted and the unsubtracted spectra appear quite
discrepant given the similarities in the light curves. The
disagreement arises solely due to two data points in each light curve
with large uncertainties, reflecting the marginal conditions under
which the observations were obtained.  Removal of those two data
points from each light curve modifies the $F_{\rm var}$ value for the
continuum-subtracted spectra only slightly, increasing from $0.075 \pm
0.011$ to $0.080 \pm 0.010$.  The $F_{\rm var}$ value for the
unsubtracted spectra, however, decreases significantly from $0.132 \pm
0.018$ to $0.099 \pm 0.012$, bringing the $F_{\rm var}$ values for the
two light curves into better agreement.

\begin{deluxetable*}{lccccccc}
\tablecolumns{8}
\tablewidth{0pt}
\tabletypesize{\footnotesize}
\tablecaption{Light-Curve Statistics}
\tablehead{
\colhead{Time Series} &
\colhead{$N$} &
\colhead{$\langle T \rangle$} &
\colhead{$T_{\rm median}$} &
\colhead{$\langle F \rangle$\tablenotemark{a}} &
\colhead{$\langle \sigma_F/F \rangle$} &
\colhead{$F_{\rm var}$} &
\colhead{$R_{\rm max}$}\\
\colhead{} &
\colhead{} &
\colhead{(days)} &
\colhead{(days)} &
\colhead{} &
\colhead{} &
\colhead{} &
\colhead{}
}
\startdata
V                 & 67 & $1.75 \pm\ 2.28$ & 1.00 & $1.35 \pm\ 0.09$ & 0.008 & $0.066 \pm\ 0.006$ & $1.336 \pm\ 0.018$ \\
H$\beta$, non-CS  & 35 & $2.68 \pm\ 1.26$ & 2.02 & $6.10 \pm\ 0.85$ & 0.023 & $0.132 \pm\ 0.018$ & $2.306 \pm\ 1.053$ \\
H$\beta$, CS      & 35 & $2.68 \pm\ 1.26$ & 2.02 & $6.68 \pm\ 0.57$ & 0.014 & $0.075 \pm\ 0.011$ & $1.459 \pm\ 0.032$ \\
\label{tab:lcstats}
\enddata 

\tablenotetext{a}{$V-$band flux density is in units of
                  $10^{-15}$\,ergs\,s$^{-1}$\,cm$^{-2}$\,\AA$^{-1}$
                  and H$\beta$ flux is in units of
                  $10^{-14}$\,ergs\,s$^{-1}$\,cm$^{-2}$.}

\end{deluxetable*}

To determine the mean time delay of the H$\beta$ emission line
relative to the continuum variations, we cross correlated the H$\beta$
light curve derived from the continuum-subtracted spectra
(Figure~\ref{fig:lc}, black points) with the $V-$band light curve
(both tabulated in Table~\ref{tab:lcdata}).  We employed the
interpolated cross-correlation function (ICCF) method
\citep{gaskell86,gaskell87} with the modifications of \citet{white94}.
This method determines the cross-correlation function (CCF) twice, by
first interpolating the continuum light curve and then by
interpolating the emission-line light curve in the second pass.  The
resultant CCF, which is the average of the two, is shown by the black
solid line in the bottom right panel of Figure~\ref{fig:lc}.  For
reference, we calculated the autocorrelation function of the $V-$band
light curve, displayed by the solid line in the top right panel of
Figure~\ref{fig:lc}.  Also displayed in Figure~\ref{fig:lc} is the CCF
for the H$\beta$ light curve derived from the original, scaled spectra
(red points) compared to the $V-$band ({\it red line}).  As expected
given the nearly identical variations in the H$\beta$ light curves,
the cross correlation functions of the two relative to $V$ are also
nearly identical.  However, the slightly reduced noise in the H$\beta$
light curve derived from the continuum-subtracted spectra provides a
higher correlation coefficient at the preferred time delay.  We
therefore focus the remainder of our analysis on the H$\beta$ light
curve derived from the continuum-subtracted spectra.

CCFs can be characterized by their maximum value ($r_{\rm max}$), the
time delay at which the CCF maximum occurs ($\tau_{\rm peak}$), and
the centroid of the points near the peak ($\tau_{\rm cent}$) above a
threshold value of $0.8 r_{\rm max}$.  However, a single CCF does not
provide any information on the uncertainties inherent in these
measurements.  We therefore employ the ``flux randomization/random
subset sampling'' (FR/RSS) method of \citet{peterson98b,peterson04},
which is a Monte Carlo approach for determining the uncertainties in
our measured time delays.  For a sample of $N$ data points, a
selection of $N$ points is chosen without regard to whether a datum
was previously chosen or not.  The typical number of points that is
not sampled in a single realization is $\sim 1/e$.  A point that is
sampled $1 \leq n \leq N$ times has its uncertainty reduced by a
factor of $n^{1/2}$.  This ``random subset sampling'' step is
therefore able to assess the uncertainty in the time delay that arises
from an individual data point in the light curve.  The ``flux
randomization'' step takes each of the selected points and modifies
the flux value by a Gaussian deviation of the uncertainty.  In this
way, the effect of the measurement uncertainties on the recovered time
delay is also assessed.  The final modified light curves are then
cross correlated with the ICCF method described above, and the values
of $r_{\rm max}$, $\tau_{\rm peak}$, and $\tau_{\rm cent}$ are
recorded.  The entire process is repeated 1000 times, and
distributions of these values are built up from all of the
realizations.  We take the medians of the cross-correlation peak
distribution (CCPD) and the cross-correlation centroid distribution
(CCCD) as $\tau_{\rm peak}$ and $\tau_{\rm cent}$, respectively.  The
uncertainties on these values are quoted so that 15.87\% of the
realizations fall above and 15.87\% fall below the range of
uncertainties, corresponding to $\pm 1 \sigma$ for a Gaussian
distribution.  The final measurements are quoted in
Table~\ref{tab:lagwidth} in both the observer's frame and the rest
frame of the AGN, and the CCCD is displayed in the bottom right panel
of Figure~\ref{fig:lc} as the blue histogram (arbitrarily scaled).
The mean of the distribution agrees well with the time delay inferred
from the CCF.

\begin{figure}
\epsscale{1.15}
\plotone{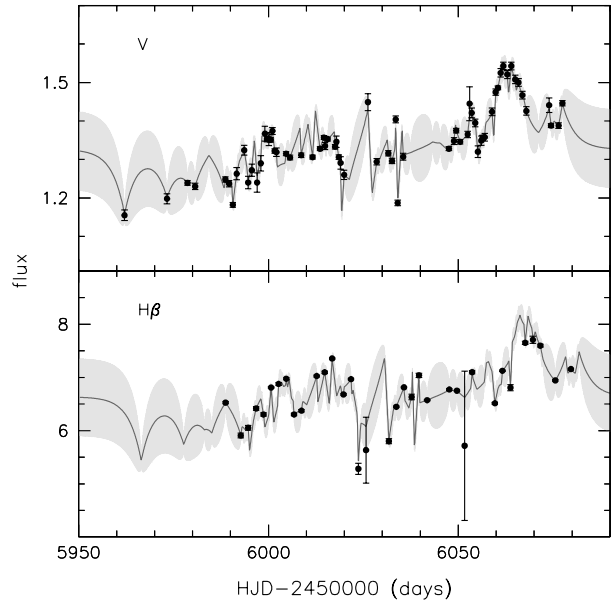}
\caption{$V-$band light curve ({\it data points, top}) and H$\beta$
  light curve derived from the continuum-subtracted spectra ({\it data
    points, bottom}).  The mean {\tt JAVELIN} models and uncertainties
  are overplotted as the solid lines and gray shaded regions,
  respectively.  The model uncertainties are derived from the standard
  deviation of the individual realizations.  {\tt JAVELIN} finds a
  best-fit H$\beta$ time delay of $4.4 \pm 0.1$\,days.}
\label{fig:javlc}
\end{figure}

For comparison, we also determined the H$\beta$ time delay using the
{\tt JAVELIN} code \citep{zu11}.  {\tt JAVELIN} employs a damped
random walk to model the continuum variations, and then determines the
best reprocessing model by quantifying the shifting and smoothing
parameters necessary to reproduce the emission-line light curve (see
Figure~\ref{fig:javlc}).  The uncertainties on the model parameters
are assessed through a Bayesian Markov Chain Monte Carlo method.  We
include the {\tt JAVELIN} time delay as $\tau_{\rm jav}$ in
Table~\ref{tab:lagwidth}.

\begin{figure}
\epsscale{1.15}
\plotone{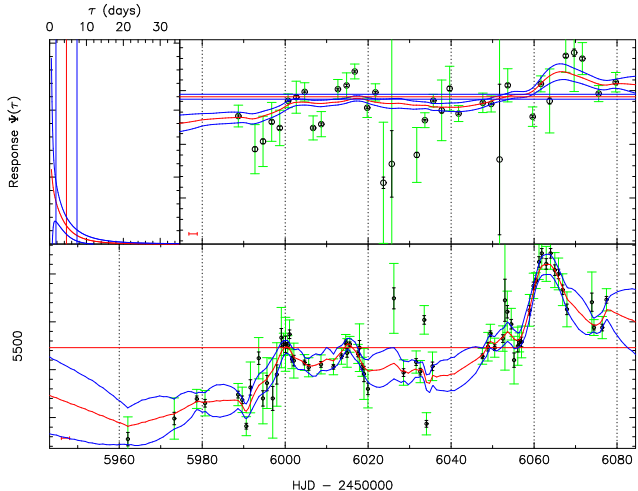}
\caption{Results of the MCMCRev fit of a linearised echo model to the
  H$\beta$ (top) and V-band (bottom) lightcurve data.  The red and
  blue curves give the mean $\pm$ rms over MCMC samples of the model
  lightcurves and the delay distribution (top left).  The mean and rms
  of MCMC samples give the mean H$\beta$ delay as $\langle \tau \rangle =4.6 \pm
  2.8$\,days.}
\label{fig:mcmcrev}
\end{figure}

Additionally, we used a Markov Chain Monte Carlo code {\tt MCMCRev} to
fit a linearised echo model to the $V-$band and H$\beta$ lightcurves
 (see Figure~\ref{fig:mcmcrev}).  This models the $V-$band lightcurve
with a fourier series constrained by the lightcurve data and with a
random-walk prior that mimics typical AGN continuum variations on
$1-100$\,d timescales.  H$\beta$ variations are modeled as an echo of
those in the $V$ band. A two-parameter delay distribution,
specifically
\begin{equation}
\Psi(\tau) = \frac{ \Psi_0 }{ 2\,\tau_0 }
	\frac{(\tau/\tau_0)^{9/4}}
	{\cosh{(\tau/\tau_0)}-1}
\ ,
\end{equation}
\noindent enforces causality ($\tau>0$) and has a width proportional
to the mean delay $\langle \tau \rangle \approx 5\,\tau_0$.  Three
further parameters are the mean H$\beta$ flux, and two factors that
scale the nominal H$\beta$ and V-band error bars.  The mean and rms of
the MCMC samples give the H$\beta$ delay as $\langle \tau \rangle =4.6 \pm
2.8$\,days, in agreement with the other techniques.

\begin{deluxetable*}{lccclcc}
\tablecolumns{7}
\tablewidth{0pt}
\tablecaption{Time Lags and Line Widths}
\tablehead{
\colhead{Frame} &
\colhead{$\tau_{\rm cent}$} &
\colhead{$\tau_{\rm peak}$} &
\colhead{$\tau_{\rm jav}$} &
\colhead{Spectrum} &
\colhead{FWHM} &
\colhead{$\sigma_{\rm line}$}
\\
\colhead{} &
\colhead{(days)} &
\colhead{(days)} &
\colhead{(days)} &
\colhead{} &
\colhead{(km\,s$^{-1}$)} &
\colhead{(km\,s$^{-1}$)} 
}
\startdata
Observed   & $5.37^{+1.87}_{-1.76}$ & $4.40^{+3.10}_{-0.80}$ & $4.40^{+0.08}_{-0.10}$ &  Mean  & $1958.4 \pm\ 74.8$   & $975.6 \pm\ 8.4$   \\
Rest-frame & $5.33^{+1.86}_{-1.75}$ & $4.37^{+3.08}_{-0.79}$ & $4.37^{+0.08}_{-0.10}$ &  RMS   & $1422.0 \pm\ 416.4$  & $664.5 \pm\ 87.3$  \\
\label{tab:lagwidth}
\enddata 
\end{deluxetable*}

\section{Line Width Measurements}

The broad emission lines in AGN spectra are interpreted as being
Doppler broadened through the bulk gas motions deep in the potential
well of the black hole.  Therefore, the width of the broad line is a
constraint on the line-of-sight velocity of the gas.  The narrow
emission lines, however, are produced by gas that is well outside the
nucleus of the AGN and does not reverberate on the time scales of a
few months.  It is therefore important that we remove the narrow
contribution to the H$\beta$ emission line before attempting to
measure the line width.

We accomplish this by using the [\ion{O}{3}] $\lambda$5007 emission
line as a template for the narrow emission lines in the spectrum.  The
template is shifted and scaled by an appropriate amount to account for
both the [\ion{O}{3}] $\lambda$4959 line and the H$\beta$ narrow
line. We adopted a scale factor of $\lambda 4959/\lambda 5007 = 0.34$
\citep{storey00} and, through trial and error, determined a scale
factor of H$\beta$/$\lambda 5007 = 0.10$.  The original and
narrow-line subtracted spectra are displayed in Figure~\ref{fig:meanspec}.

From the narrow-line subtracted spectra, we determined the emission
line width in both the mean and rms of the continuum-subtracted
spectra.  A local linear continuum was determined from two continuum
windows on either side of the emission line, and the width was
determined directly from the measurements above this local continuum.
We report the line width as the full width at half the maximum flux
(FWHM) and also as the second moment of the line profile, or the line
dispersion, $\sigma_{\rm line}$.

The uncertainties on the line width measurements were determined from
a Monte Carlo random subset sampling method.  For a set of $N$
spectra, we select $N$ without regard to whether a spectrum was
previously chosen or not.  The mean and rms of this subset are
determined, and the FWHM and $\sigma_{\rm line}$ are tabulated.  The
process is repeated 1000 times, and a distribution of each measurement
is built up.  In this way, the effect of any particular spectrum on
the line width measurements is assessed.  We also included a slight
modification in which the continuum windows on either side of the
emission line were allowed to vary in size and exact placement within
an acceptable range, thereby assessing the effect of the continuum
window choice on the final measurements.  This modification generally
has little or no effect on the line widths derived from the rms
spectrum, where noise already dominates the uncertainties, but
slightly increases the uncertainties on the line widths derived from
the mean spectrum \citep{bentz09c}.  The mean and standard deviation
of each distribution are adopted as the measurement value and its
uncertainty, respectively.

Finally, we also corrected for the dispersion of the spectrograph
following the method employed by \citet{peterson04}, in which the
observed line width can be described as a combination of the intrinsic
line width, $\Delta \lambda_{\rm true}$, and the spectrograph
dispersion, $\Delta \lambda_{\rm disp}$, such that
\begin{equation}
\Delta \lambda_{\rm obs}^2 \approx \Delta \lambda_{\rm true}^2 + \Delta \lambda_{\rm disp}^2.
\end{equation}
\noindent In this case, it is not possible to measure $\Delta
\lambda_{\rm disp}$ from sky lines or arc lamps employed for
wavelength calibration, because in both of those cases, the source
fills the entire slit.  However, the angular size of the unresolved
AGN point source is set by the seeing, which varies throughout the
campaign but is almost always smaller than the 5\arcsec\ width of the
slit.  Our typical approach is to therefore search the literature for
very high resolution measurements of the width of the [\ion{O}{3}]
lines, to serve as a measurement of $\Delta \lambda_{\rm true}$,
allowing $\Delta \lambda_{\rm disp}$ to be determined.  Such
measurements do not exist for MCG-06-30-15, but they do exist for
NGC\,1566, another Seyfert galaxy that we have monitored with the same
instrument and setup.

For NGC\,1566, \citet{whittle92} measured ${\rm FWHM} = 280$\,km\,s$^{-1}$
for [\ion{O}{3}] $\lambda 5007$ through a small slit, with a high
resolution, and under good observing conditions.  From our own spectra
of NGC\,1566 taken with RCSpec on the SMARTS 1.5-m telescope, we
determined ${\rm FWHM} = 8.26$\,\AA\ for [\ion{O}{3}] $\lambda 5007$.  We
therefore deduce a value of $\Delta \lambda_{\rm disp} =
6.8$\,\AA\ and adopt this value for our observations of MCG-06-30-15.
The final dispersion-corrected line widths and uncertainties for the
mean and rms H$\beta$ broad line profiles are tabulated in
Table~\ref{tab:lagwidth}.

\section{Black Hole Mass}

The black hole mass is generally derived from reverberation-mapping
measurements as
\begin{equation}
M_{BH} = f \frac{RV^2}{G}
\end{equation}
where $R$ is taken to be $c\tau$, the speed of light times the mean
time delay of a broad emission line relative to continuum variations,
$V$ is the line-of-sight velocity of the gas in the broad line region
and is determined from the emission line width, and $G$ is the
gravitational constant.  

The factor $f$ is a scaling factor that accounts for the detailed
geometry and kinematics of the gas in the broad line region, which is
generally unknown.  In practice, it has become common to determine the
population average multiplicative factor, $\langle f \rangle$,
necessary to bring the \msigma\ relationship for AGNs with
reverberation masses into agreement with the \msigma\ relationship for
nearby galaxies with dynamical black hole masses (e.g.,
\citealt{gultekin09,mcconnell13,kormendy13}).  In this way, the overall scale for
reverberation masses should be unbiased, but the mass of any single
AGN is expected to be uncertain by a factor of 2-3.  The value of
$\langle f \rangle$ has varied in the literature from 5.5
\citep{onken04} to 2.8 \citep{graham11}, depending on which objects
are included and the specifics of the measurements.  We adopt the
value determined by \citet{grier13} of $\langle f \rangle = 4.3 \pm
1.1$.

Our preferred combination of measurements is $\tau_{\rm cent}$ for the
time delay and $\sigma_{\rm line}$ measured from the rms spectrum for
the line width.  Combined with our adopted value of $\langle f
\rangle$, we determine a black hole mass of $(1.6\pm0.4) \times
10^6$\,M$_{\odot}$ for MCG-06-30-15.

\section{Discussion}


We present here the first optical emission-line reverberation results
for MCG-06-30-15, but the well-studied nature of this AGN ensures that
we have ample comparisons available in the literature with which we
can assess our results.  \citet{lira15} describe a long-term
monitoring campaign in X-ray, optical, and near-IR bands from which
several broad-band time delays were measured.  In particular, they
find that the near-IR bands lag the $B$ and $V$ bands by 13, 20, and
26\,days in $J$, $H$, and $K$ respectively.  While our monitoring
campaign was not contemporaneous with that described by
\citet{lira15}, it was carried out the following observing season.
Furthermore, \citet{kara14} find that the luminosity state of
MCG-06-30-15 did not change significantly over the period between 2001
and 2013, and the light curve from the Swift/BAT hard X-ray transient
monitor shows no luminosity state changes between 2005 and 2015
\citep{krimm13}.  Comparison of our measured H$\beta$ time delay of
$\tau_{\rm cent} = 5.3 \pm\ 1.8$\,days to the near-IR delays places
the inner edge of the dust torus outside the BLR, as has been found
for other Seyferts \citep{clavel89,suganuma06,koshida14}.
Furthermore, our H$\beta$ time delay compares remarkably well with the
findings of \citet{koshida14} that $\tau(K) / \tau({\rm H}\beta)
\approx 4-5$.  These findings are also in keeping with the scenario
proposed by \citet{netzer93} in which the dust torus creates the outer
edge of the BLR through suppression of line emission by the dust
grains.

\subsection{AGN Radius$-$Luminosity Relationship}

The empirical relationship between the AGN BLR radius and the AGN
optical luminosity \citep{kaspi00,kaspi05,bentz06a,bentz09b,bentz13} is
a well-known scaling relationship derived from the set of
reverberation mapping measurements for relatively nearby AGNs.  The
calibrated relationship relies on H$\beta$ reverberation results and
measurements of the continuum luminosity at 5100\,\AA, and it provides
a quick way to estimate black hole masses without investing in time-
and resource-intensive reverberation mapping programs for every target
of interest.  The \rl\ relationship has been found to be in good
agreement with simple expectations from photoionization physics, once
the luminosity measurements were corrected for the host-galaxy
starlight contribution measured through the reverberation-mapping
spectroscopic aperture \citep{bentz06a,bentz09b,bentz13}.  The scatter
has also been found to be quite low, $<0.2$\,dex \citep{bentz13},
implying that AGNs are mostly luminosity-scaled versions of each
other.

Starlight corrections are especially important for nearby AGNs, like
MCG-06-30-15, because they can provide a significant fraction of the
flux through the reverberation-mapping spectroscopic aperture.  These
corrections are generally obtained through two-dimensional surface
brightness modeling of high-resolution AGN host galaxy images.  The
decomposition allows the AGN PSF to be accurately separated from the
host-galaxy and the underlying sky, and thus an ``AGN free'' image can
be recovered from which the starlight flux can be measured.
MCG-06-30-15 was observed with \hst\ and the UVIS channel of WFC3
through the F547M filter as part of program GO-11662 to image the host
galaxies of the LAMP 2008 AGN sample \citep{bentz13}.  A single orbit
was split into two pointings separated by a small angle maneuver, and
at each pointing a set of three exposures was taken, each exposure
graduated in exposure time (short, medium, and long).  The saturated
pixels in the AGN core in the long exposures are corrected by scaling
up the same pixels from the shorter, unsaturated exposures by the
ratio of the exposure times.  In this way, the graduated exposure
times allow the dynamic range of the final drizzled image to
significantly exceed the dynamic range of the detector itself.  The
total exposure time of the final combined, drizzled image is 2290\,s.

\begin{figure*}
\epsscale{1.15}
\plotone{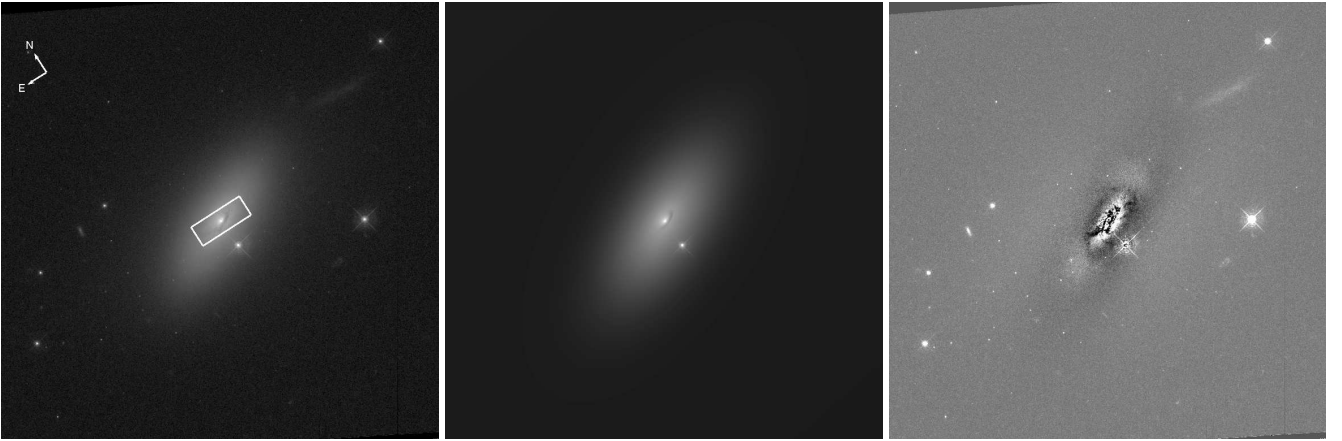}
\caption{{\it Left:} {\it HST} image of the host-galaxy of
  MCG-06-30-15, displayed with a logarithmic stretch.  The scale of is
  80\arcsec\,$\times$80\arcsec and is only a portion of the full field
  of view.  The white rectangle centered on the galaxy nucleus shows
  the ground-based spectroscopic monitoring aperture. {\it Middle:}
  {\tt GALFIT} model, displayed with the same stretch as the image. {\it
    Right:} Residuals after subtracting the model from the image,
  displayed with a linear stretch centered around zero counts.}
\label{fig:galfit}
\end{figure*}

Two-dimensional surface brightness fitting of the \hst\ image was
carried out with the {\tt GALFIT} software \citep{peng02,peng10}.  We
fit the host-galaxy of MCG-06-30-15 with a \sersic\ bulge and an
exponential disk with an inner radial (truncation) function to
approximate the dust lane.  A single Fourier mode ($m=1$) was also
allowed for each of these components, to account for gross
perturbations on the initial parametric models.  The AGN and nearby
star were fit with a model PSF generated by the {\tt Starfit}
algorithm \citep{hamilton14}, which starts with a TinyTim PSF model
\citep{krist93} and then fits the subpixel centering and the telescope
focus.  The underlying sky background was also fit as a gradient, and
we used the entire field of view provided by WFC3 to ensure that it
was properly constrained, even though the galaxy itself only covers a
small portion of the UVIS1 camera.  The parameters for our best-fit
model are tabulated in Table~\ref{tab:galfit}, and
Figure~\ref{fig:galfit} displays a region of the \hst\ image centered
on the galaxy ({\it left}), the best-fit model image ({\it center}),
and the residuals after subtraction of the model from the image ({\it
  right}).

\begin{deluxetable*}{lclcccccccl}
\tablecolumns{10}
\tablewidth{0pt}
\tabletypesize{\scriptsize}
\tablecaption{Surface Brightness Decomposition}
\tablehead{
\colhead{\#} &
\colhead{PSF+sky} &
\colhead{$\Delta x$ (\arcsec)} &
\colhead{$\Delta y$ (\arcsec)} &
\colhead{$m_{\rm stmag}$\tablenotemark{a}} &
\colhead{...} &
\colhead{Sky (cts)} &
\colhead{$\frac{d{\rm sky}}{dx}$ ($10^{-4}$ cts)} &
\colhead{$\frac{d{\rm sky}}{dy}$ ($10^{-4}$ cts)} &
\colhead{Note} \\
\colhead{} &
\colhead{sersic} &
\colhead{$\Delta x$ (\arcsec)} &
\colhead{$\Delta y$ (\arcsec)} &
\colhead{$m_{\rm stmag}$} &
\colhead{$r_{\rm e}$ (\arcsec)} &
\colhead{$n$} &
\colhead{$b/a$} &
\colhead{PA (deg)} &
\colhead{} \\
\colhead{} &
\colhead{sersic3} &
\colhead{$\Delta x$ (\arcsec)} &
\colhead{$\Delta y$ (\arcsec)} &
\colhead{$\Sigma_{\rm stmag}$} &
\colhead{$r_{\rm e}$ (\arcsec)} &
\colhead{$n$} &
\colhead{$b/a$} &
\colhead{PA (deg)} &
\colhead{} \\
\colhead{} &
\colhead{radial} &
\colhead{$\Delta x$ (\arcsec)} &
\colhead{$\Delta y$ (\arcsec)} &
\colhead{\nodata} &
\colhead{$r_{\rm break}$ (\arcsec)} &
\colhead{$\Delta r_{\rm soft}$ (\arcsec)} &
\colhead{$b/a$} &
\colhead{PA (deg)} &
\colhead{} \\
\colhead{} &
\colhead{fourier} &
\colhead{\nodata} &
\multicolumn{2}{c}{mode: $a_{\rm m}$ , $\phi$ (deg)} &
\colhead{\nodata} &
\colhead{\nodata} &
\colhead{\nodata} &
\colhead{\nodata} &
\colhead{} \\
\colhead{(1)} &
\colhead{(2)} &
\colhead{(3)} &
\colhead{(4)} &
\colhead{(5)} &
\colhead{(6)} &
\colhead{(7)} &
\colhead{(8)} &
\colhead{(9)} &
\colhead{(10)} 
}
\startdata
1,2 & PSF+sky        & 0.000     & 0.000    & 16.26  & \nodata  & 31.72  & -3.7	 & 4.2     &   \\
3   & sersic 	     &	0.038	 & -0.008   & 15.70  &	1.014	& 1.9	 & 0.47	 & -36.8   & bulge  \\
    & ~~fourier	     & 1: -0.418 & -96.2    &	     &		&	 & 	 &	   &   \\
4   & sersic3	     &	0.132	 & -0.092   & 18.66  &	8.216	& [1.0]	 & 0.48	 & -32.1   & disk  \\
    & ~~radial,inner &	1.036	 & 0.764    &\nodata &	2.149	& 2.195	 & 0.35	 & -24.7   & dust lane  \\
    & ~~fourier	     & 1: 0.722	 & 116.5    &	     &		&	 &	 &	   &   \\
\multicolumn{2}{l}{merit} & \multicolumn{3}{l}{$\chi^2=645340672.0$} & \multicolumn{2}{l}{$N_{\rm dof}=15783039$}
                          &  $N_{\rm free}$=29       & $\chi^2_{\nu}=40.88$     &                \\ 
\label{tab:galfit}
\enddata  
\tablecomments{Values in square brackets were held fixed during the surface brightness model fitting.}
\tablenotetext{a}{The STmag magnitude system is based on the absolute physical flux per unit wavelength.}
\end{deluxetable*}

Using our best-fit model, we created a sky- and AGN-subtracted image
of MCG-06-30-15.  From this image, we measured the host-galaxy flux
density through the ground-based spectroscopic monitoring aperture
(depicted as the white rectangle in the left panel of
Figure~\ref{fig:galfit}). The scaling factor necessary to correct the
flux density from the effective wavelength of the \hst\ filter to
5100$\times$(1+z) was determined with {\it synphot} and a template
galaxy bulge spectrum \citep{kinney96}.  Our determination of the
host-galaxy flux density at 5100$\times$(1+z) is $f_{\rm gal} = (4.28
\pm 0.43) \times 10^{-15}$\,erg\,s$^{-1}$\,cm$^{-2}$\,\AA$^{-1}$.
The average flux density at 5100$\times$(1+z) was determined from our
scaled spectra to be $f_{\rm obs} = (5.23 \pm 0.08) \times
10^{-15}$\,erg\,s$^{-1}$\,cm$^{-2}$\,\AA$^{-1}$.  Correcting for the
host-galaxy contribution, we deduce an AGN-only flux density of
$f_{\rm AGN} = (0.95 \pm 0.49) \times
10^{-15}$\,erg\,s$^{-1}$\,cm$^{-2}$\,\AA$^{-1}$.

Unfortunately, the distance to MCG-06-30-15 is not particularly well
constrained.  The luminosity distance implied by the galaxy redshift
is $D_L = 32.5$\,Mpc.  However, the Extragalactic Distance Database
\citep{tully09} reports $D = 25.5 \pm 3.5$ from their cosmic flows
model and the group membership of MCG-06-30-15 \citep{tully13}.  Taken
at face value, this $\sim 30$\% disagreement in distance leads to a
factor of 1.6 uncertainty in the luminosity.  Additionally, there are
only three galaxies contributing to the group distance determination,
and the individual distance estimates for these three galaxies range
from $20-36$\,Mpc.  As part of a separate program to determine
\citet{tully77} distances to AGN host galaxies, we observed
MCG-06-30-15 with the Green Bank Telescope, but we were unable to
detect \ion{H}{1} 21\,cm emission with 3.5\,hrs of on-source time.
For our purposes here, we adopt the cosmic flows estimate and its
uncertainty, but we note that it will be important to better constrain
the distance to this galaxy in order to determine more accurate
physical parameters (including, but not limited to, any luminosity
measurements).  After correcting for Galactic extinction along the
line of sight as determined by \citet{schlafly11}, we find $\log
\lambda L_{\lambda} (5100\,{\rm \AA}) = 41.65 \pm 0.25$\,erg\,s$^{-1}$.

\begin{figure}[h!]
\epsscale{1.15}
\plotone{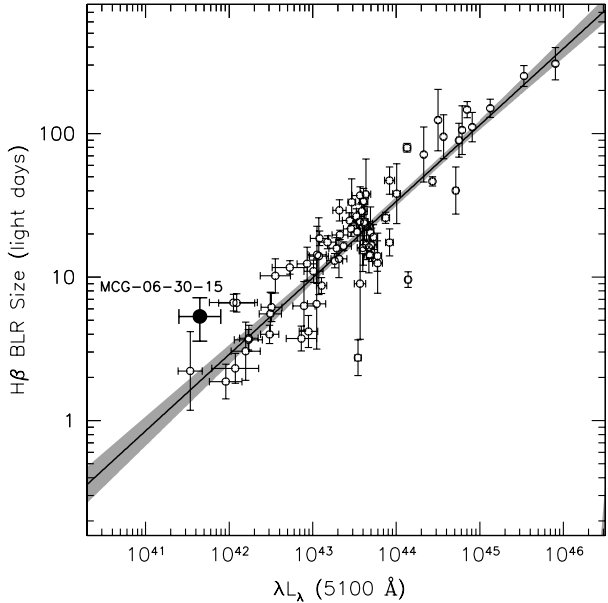}
\caption{The radius$-$luminosity relationship for AGNs (open points
  and fit; \citealt{bentz13}) with the H$\beta$ time delay and AGN
  luminosity for MCG-06-30-15 from this work plotted as the filled
  point.}
\label{fig:rl}
\end{figure}

Figure~\ref{fig:rl} depicts the location of MCG-06-30-15 on the
\rl\ relationship.  We have not determined a new best fit to the
relationship, but have simply recreated the plot from \citet{bentz13}.
MCG-06-30-15 is fairly consistent with the typical scatter around the
relationship.  We note that if we were to adopt one of the other time
delay measurements (such as $\tau_{\rm jav}$), or the luminosity
distance from the galaxy redshift, the agreement would be even better.
With the adopted assumptions, and furthermore assuming that $L_{\rm
  bol} = 4.9 + 0.9 \lambda L_{\lambda}(5100\,{\rm \AA})$
\citep{runnoe12}, we estimate $L/L_{\rm Edd} = 0.04$.

\subsection{Black Hole Mass Consistency}

Our measurement of $M_{\rm BH} = (1.6 \pm 0.4) \times
10^6$\,M$_{\odot}$ for MCG-06-30-15 is in excellent agreement with the
value determined by \citet{mchardy05} of $M_{\rm BH} =
2.9^{+1.8}_{-1.6}\times10^6$\,M$_{\odot}$.  Their work assumed a
linear scaling between \mbh\ and the X-ray power spectral density
break, with the relationship anchored to the measurements for the
Galactic black hole Cygnus X-1.  This agreement therefore bolsters the
claim that supermassive black holes are simply analogs of Galactic
black holes, but scaled up in mass (e.g., \citealt{mchardy06}).

\citet{raimundo13} describe VLT SINFONI integral field spectroscopic
observations of the innermost $\sim 0.5$\,kpc of the galaxy in the $H$
band.  Although their observations were somewhat shallow (total
on-source exposure time of 1.3 hours), they attempted to constrain the
black hole mass with the Jeans Anisotropic Model method
\citep{cappellari08}.  Intriguingly, they find a best-fit value of
$M_{\rm BH}=4\times10^6$\,M$_{\odot}$ (assuming $D=37$\,Mpc), although
they caution that there is actually a stronger constraint on an upper
limit of $M_{\rm BH} < 6 \times 10^7$\,M$_{\odot}$ than on the
best-fit mass.

One of our original reasons for targeting MCG-06-30-15 included the
fact that it might be possible to determine a black hole mass through
both reverberation mapping and stellar dynamical modeling for this
nearby AGN.  The sample of objects for which we are able to compare
these two mass determination methods is extremely small for two
reasons: (1) stellar dynamical modeling is limited by spatial
resolution, and therefore distance; and (2) broad-lined AGNs in the
local Universe are quite rare, and therefore generally far away.
Only two galaxies have published masses from both
methods thus far --- NGC\,4151 \citep{bentz06a,onken14} and NGC\,3227
\citep{davies06,denney10}.

A useful metric for determining whether a stellar dynamical mass is
likely to be achievable is to determine whether the black hole sphere
of influence ($r_h$) could be resolved with the observations, where
\begin{equation}
  r_h = \frac{GM_{BH}}{\sigma_{\star}^2}.
\end{equation}
\noindent Combining our mass with the value of
$\sigma_{\star}=109$\,km\,s$^{-1}$ determined by \citet{raimundo13}
and the distance of 25.5\,Mpc adopted above, we estimate
$r_h=0.005$\arcsec.  This scale is not resolvable with
currently-available instruments, although \citet{gultekin09} argue that
it is not strictly necessary to resolve $r_h$ to obtain a useful
constraint on $M_{\rm BH}$.  Furthermore, the best-fit black hole mass
derived by \citet{raimundo13}, even with shallow observations and a
spatial scale of 0.05\arcsec, suggests that it could be worthwhile to
pursue a stellar dynamical mass constraint for MCG-06-30-15.  In this
case, an accurate distance will be even more necessary, as dynamical
masses scale linearly with the assumed distance.

Time lags between different X-ray energy bands have also been detected
in MCG-06-30-15 \citep{emman11, demarco13, kara14}. Of particular
interest are soft X-ray lags (where low energy X-rays lag behind
higher energy X-rays), likely due to X-ray reverberation
\citep{fabian09}. \citet{emman11} first detected a soft lag of
approximately 20\,s in MCG-06-30-15.  A systematic search for, and
analysis of, soft lags in X-ray variable AGN found that the amplitude
of the soft lags and Fourier frequency where they are observed scales
with black hole mass \citep{demarco13}.  MCG-06-30-15 is one of the 15
soft lag detections used to determine the scaling relation, with a
black hole mass estimated from the \rl\ relationship.  Ignoring that
MCG-06-30-15 was used in determined the soft lag scaling relation, and
that the scaling relation is subject to selection biases
\citep{demarco13}, we can use the soft lag to estimate the black hole
mass.  \citet{demarco13} measure a soft lag of $26.4\pm12.7$ s, which
predicts a black hole mass of $1.1\times10^6$\,M$_{\odot}$, consistent
with the reverberation mass we have determined in this work.

\begin{figure}[h!]
\epsscale{1.15}
\plotone{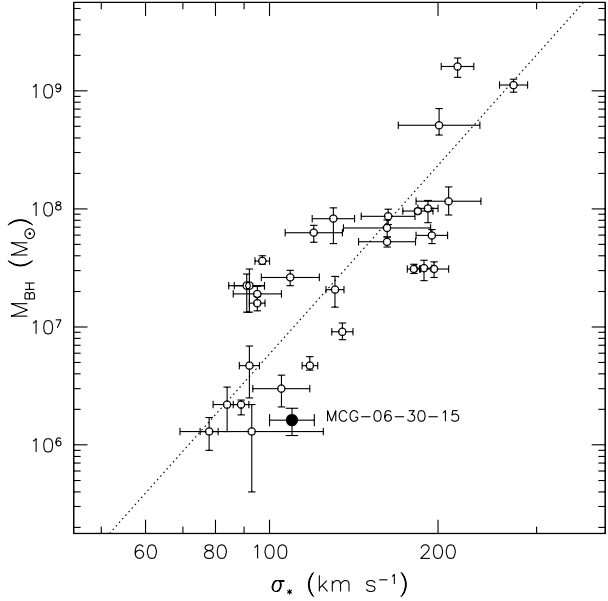}
\caption{The $M_{\rm BH}-\sigma_{\star}$ relationship for AGNs (open
  points and fit; \citealt{grier13}) with MCG-06-30-15 plotted as the
  filled point.}
\label{fig:msig}
\end{figure}

\subsection{AGN $M-\sigma$ Relationship}

We also examine the black hole mass we have derived for MCG-06-30-15
in light of the \msigma\ relationship for other AGNs with
reverberation masses.  \citet{raimundo13} contrained the bulge stellar
velocity dispersion from their VLT SINFONI velocity dispersion maps
using a pseudoslit geometry and determined $\sigma_{\star} =
109$\,km\,s$^{-1}$.  This value is somewhat larger than the value of
$\sigma_{\star} = 93.5 \pm 8.5$\,km\,s$^{-1}$ determined by
\citet{mchardy06} from longslit spectroscopy.

In Figure~\ref{fig:msig}, we show the AGN \msigma\ relationship from
\citet{grier13}.  MCG-06-30-15 sits a bit below and to the right of
the relationship, but appears to be fairly consistent within the
scatter.  Adoption of the \citet{mchardy06} value of $\sigma_{\star}$
would further bolster the agreement.

We note that we have an independent project currently in progress that
will recalibrate the AGN \msigma\ relationship using velocity
dispersions derived solely from integral field spectroscopy, which
will be important for removing any rotational broadening effects from
the $\sigma_{\star}$ measurements among the rest of the sample
\citep{batiste16}, as well as any scatter imposed by the selection of
a specific position angle for longslit observations.  We intend to
revisit the location of MCG-06-30-15 on this relationship at that
time.

\section{Summary}

We have determined a reverberation time delay for the broad H$\beta$
emission line in the spectrum of MGC-06-30-15 of $\tau_{\rm
  cent}=5.3\pm1.8$\,days in the rest-frame of the AGN.  The measured
time delay is in good agreement with the AGN \rl\ relationship.  It
also agrees with the relationship between H$\beta$ and near-IR time
delays, where the effective optical BLR size is approximately 4-5
times smaller than the inner edge of the dust torus. Combining the
H$\beta$ time delay measurement with the width of the emission line in
the variable part of the spectrum, we constrain a central black hole
mass of $M_{\rm BH} = (1.6 \pm 0.4) \times 10^6$\,M$_{\odot}$.  This
value is in good agreement with estimates from the X-ray power
spectral density break and \msigma\ relationships.

\acknowledgements 

We thank the referee for comments that improved the presentation of
this paper.  MCB gratefully acknowledges support from the NSF through
CAREER grant AST-1253702.  KH acknowledges support from UK Science and
Technology Facilities Council (STFC) grant ST/M001296/1.  We thank the
SMARTS Consortium members and CTIO staff for efforts to make
observations at the CTIO/SMARTS 1.5m telescope possible. This work
makes use of observations from the LCOGT network and is based on
observations with the NASA/ESA {\it Hubble Space Telescope}.  We are
grateful for support of this work through grant \hst\ GO-11662 from
the Space Telescope Science Institute, which is operated by the
Association of Universities for Research in Astronomy, Inc., under
NASA contract NAS5-26555. The Liverpool Telescope is operated on the
island of La Palma by Liverpool John Moores University in the Spanish
Observatorio del Roque de los Muchachos of the Instituto de
Astrofisica de Canarias with financial support from the UK Science and
Technology Facilities Council.  This research has made use of the
NASA/IPAC Extragalactic Database (NED) which is operated by the Jet
Propulsion Laboratory, California Institute of Technology, under
contract with the National Aeronautics and Space Administration and
the SIMBAD database, operated at CDS, Strasbourg, France.


\begin{thebibliography}{80}
\expandafter\ifx\csname natexlab\endcsname\relax\def\natexlab#1{#1}\fi

\bibitem[{{Alard}(2000)}]{alard00}
{Alard}, C. 2000, \aaps, 144, 363

\bibitem[{{Alard} \& {Lupton}(1998)}]{alard98}
{Alard}, C., \& {Lupton}, R.~H. 1998, \apj, 503, 325

\bibitem[{{Antonucci}(1993)}]{antonucci93}
{Antonucci}, R. 1993, \araa, 31, 473

\bibitem[{{Batiste} \& {Bentz}(2016)}]{batiste16}
{Batiste}, M., \& {Bentz}, M.~C. 2016, in American Astronomical Society Meeting
  Abstracts, Vol. 227, American Astronomical Society Meeting Abstracts, 104.07

\bibitem[{{Bentz} {et~al.}(2013){Bentz}, {Denney}, {Grier}, 
  {et~al.}}]{bentz13}
{Bentz}, M.~C., {Denney}, K.~D., {Grier}, C.~J., et~al. 2013, \apj,
  767, 149

\bibitem[{{Bentz} {et~al.}(2009{\natexlab{a}}){Bentz}, {Peterson}, {Netzer},
  {Pogge}, \& {Vestergaard}}]{bentz09b}
{Bentz}, M.~C., {Peterson}, B.~M., {Netzer}, H., {Pogge}, R.~W., \&
  {Vestergaard}, M. 2009{\natexlab{a}}, \apj, 697, 160

\bibitem[{{Bentz} {et~al.}(2006){Bentz}, {Peterson}, {Pogge}, {Vestergaard}, \&
  {Onken}}]{bentz06a}
{Bentz}, M.~C., {Peterson}, B.~M., {Pogge}, R.~W., {Vestergaard}, M., \&
  {Onken}, C.~A. 2006, \apj, 644, 133

\bibitem[{{Bentz} {et~al.}(2009{\natexlab{b}}){Bentz}, {Walsh}, {Barth},
  {et~al.}}]{bentz09c}
{Bentz}, M.~C., {Walsh}, J.~L., {Barth}, A.~J., et~al. 2009{\natexlab{b}}, \apj, 705, 199

\bibitem[{{Bentz} {et~al.}(2008)}]{bentz08}
{Bentz}, M.~C., {et~al.} 2008, \apjl, 689, L21

\bibitem[{{Blandford} \& {McKee}(1982)}]{blandford82}
{Blandford}, R.~D., \& {McKee}, C.~F. 1982, \apj, 255, 419

\bibitem[{{Brenneman} \& {Reynolds}(2006)}]{brenneman06}
{Brenneman}, L.~W., \& {Reynolds}, C.~S. 2006, \apj, 652, 1028

\bibitem[{{Cappellari}(2008)}]{cappellari08}
{Cappellari}, M. 2008, \mnras, 390, 71

\bibitem[{{Chiang} \& {Fabian}(2011)}]{chiang11}
{Chiang}, C.-Y., \& {Fabian}, A.~C. 2011, \mnras, 414, 2345

\bibitem[{{Clavel} {et~al.}(1989){Clavel}, {Wamsteker}, \& {Glass}}]{clavel89}
{Clavel}, J., {Wamsteker}, W., \& {Glass}, I.~S. 1989, \apj, 337, 236

\bibitem[{{Davies} {et~al.}(2006){Davies}, {Thomas}, {Genzel}, 
  {et~al.}}]{davies06}
{Davies}, R.~I., {Thomas}, J., {Genzel}, R., et~al. 2006, \apj, 646, 754

\bibitem[{{De Marco} {et~al.}(2013){De Marco}, {Ponti}, {Cappi}, {Dadina},
  {Uttley}, {Cackett}, {Fabian}, \& {Miniutti}}]{demarco13}
{De Marco}, B., {Ponti}, G., {Cappi}, M., {Dadina}, M., {Uttley}, P.,
  {Cackett}, E.~M., {Fabian}, A.~C., \& {Miniutti}, G. 2013, \mnras, 431, 2441

\bibitem[{{Denney} {et~al.}(2010){Denney}, {Peterson}, {Pogge}, 
  {et~al.}}]{denney10}
{Denney}, K.~D., {Peterson}, B.~M., {Pogge}, R.~W., et~al.  2010, \apj, 721, 715

\bibitem[{{Edelson} {et~al.}(2002){Edelson}, {Turner}, {Pounds}, {Vaughan},
  {Markowitz}, {Marshall}, {Dobbie}, \& {Warwick}}]{edelson02}
{Edelson}, R., {Turner}, T.~J., {Pounds}, K., {Vaughan}, S.,
  {Markowitz}, A., {Marshall}, H., {Dobbie}, P., \& {Warwick}, R. 2002, \apj, 568, 610

\bibitem[{{Emmanoulopoulos} {et~al.}(2011){Emmanoulopoulos}, {McHardy}, \&
  {Papadakis}}]{emman11}
{Emmanoulopoulos}, D., {McHardy}, I.~M., \& {Papadakis}, I.~E. 2011, \mnras,
  416, L94

\bibitem[{{Fabian}(2012)}]{fabian12}
{Fabian}, A.~C. 2012, \araa, 50, 455

\bibitem[{{Fabian} {et~al.}(2009){Fabian}, {Zoghbi}, {Ross}, 
  {et~al.}}]{fabian09}
{Fabian}, A.~C., {Zoghbi}, A., {Ross}, R.~R., et~al. 2009, \nat, 459, 540

\bibitem[{{Fath}(1913)}]{fath13}
{Fath}, E.~A. 1913, \apj, 37

\bibitem[{{Ferrarese} \& {Ford}(2005)}]{ferrarese05}
{Ferrarese}, L., \& {Ford}, H. 2005, \ssr, 116, 523

\bibitem[{{Ferrarese} \& {Merritt}(2000)}]{ferrarese00}
{Ferrarese}, L., \& {Merritt}, D. 2000, \apjl, 539, L9

\bibitem[{{Gaskell} \& {Peterson}(1987)}]{gaskell87}
{Gaskell}, C.~M., \& {Peterson}, B.~M. 1987, \apjs, 65, 1

\bibitem[{{Gaskell} \& {Sparke}(1986)}]{gaskell86}
{Gaskell}, C.~M., \& {Sparke}, L.~S. 1986, \apj, 305, 175

\bibitem[{{Gebhardt} {et~al.}(2000){Gebhardt}, {Bender}, {Bower}, 
  {et~al.}}]{gebhardt00}
{Gebhardt}, K., {Bender}, R., {Bower}, G., et~al. 2000, \apjl, 539, L13

\bibitem[{{Graham} {et~al.}(2011){Graham}, {Onken}, {Athanassoula}, \&
  {Combes}}]{graham11}
{Graham}, A.~W., {Onken}, C.~A., {Athanassoula}, E., \& {Combes}, F. 2011,
  \mnras, 412, 2211

\bibitem[{{Grier} {et~al.}(2013){Grier}, {Martini}, {Watson}, 
  {et~al.}}]{grier13}
{Grier}, C.~J., {Martini}, P., {Watson}, L.~C., et~al. 2013, \apj, 773, 90

\bibitem[{{Grier} {et~al.}(2012){Grier}, {Peterson}, {Pogge}, 
  {et~al.}}]{grier12b}
{Grier}, C.~J., {Peterson}, B.~M., {Pogge}, R.~W., et~al. 2012, \apj, 755, 60

\bibitem[{{G{\"u}ltekin} {et~al.}(2009){G{\"u}ltekin}, {Richstone}, {Gebhardt},
  {et~al.}}]{gultekin09}
{G{\"u}ltekin}, K., {Richstone}, D.~O., {Gebhardt}, K., et~al. 2009, \apj, 698, 198

\bibitem[{{Hamilton}(2014)}]{hamilton14}
{Hamilton}, T.~S. 2014, in American Astronomical Society Meeting Abstracts,
  Vol. 223, American Astronomical Society Meeting Abstracts \#223, 145.02

\bibitem[{{Heckman} \& {Best}(2014)}]{heckman14}
{Heckman}, T.~M., \& {Best}, P.~N. 2014, \araa, 52, 589

\bibitem[{{Kara} {et~al.}(2014){Kara}, {Fabian}, {Marinucci}, {Matt}, {Parker},
  {Alston}, {Brenneman}, {Cackett}, \& {Miniutti}}]{kara14}
{Kara}, E., {Fabian}, A.~C., {Marinucci}, A., {Matt}, G., {Parker}, M.~L.,
  {Alston}, W., {Brenneman}, L.~W., {Cackett}, E.~M., \& {Miniutti}, G. 2014,
  \mnras, 445, 56

\bibitem[{{Kaspi} {et~al.}(2005){Kaspi}, {Maoz}, {Netzer}, {Peterson},
  {Vestergaard}, \& {Jannuzi}}]{kaspi05}
{Kaspi}, S., {Maoz}, D., {Netzer}, H., {Peterson}, B.~M., {Vestergaard}, M., \&
  {Jannuzi}, B.~T. 2005, \apj, 629, 61

\bibitem[{{Kaspi} {et~al.}(2000){Kaspi}, {Smith}, {Netzer}, {Maoz}, {Jannuzi},
  \& {Giveon}}]{kaspi00}
{Kaspi}, S., {Smith}, P.~S., {Netzer}, H., {Maoz}, D., {Jannuzi}, B.~T., \&
  {Giveon}, U. 2000, \apj, 533, 631

\bibitem[{{King} \& {Pounds}(2015)}]{king15}
{King}, A., \& {Pounds}, K. 2015, \araa, 53, 115

\bibitem[{{Kinney} {et~al.}(1996){Kinney}, {Calzetti}, {Bohlin}, {McQuade},
  {Storchi-Bergmann}, \& {Schmitt}}]{kinney96}
{Kinney}, A.~L., {Calzetti}, D., {Bohlin}, R.~C., {McQuade}, K.,
  {Storchi-Bergmann}, T., \& {Schmitt}, H.~R. 1996, \apj, 467, 38

\bibitem[{{Koleva} {et~al.}(2009){Koleva}, {Prugniel}, {Bouchard}, \&
  {Wu}}]{koleva09}
{Koleva}, M., {Prugniel}, P., {Bouchard}, A., \& {Wu}, Y. 2009, \aap, 501, 1269

\bibitem[{{Kormendy} \& {Ho}(2013)}]{kormendy13}
{Kormendy}, J., \& {Ho}, L.~C. 2013, \araa, 51, 511

\bibitem[{{Koshida} {et~al.}(2014){Koshida}, {Minezaki}, {Yoshii}, 
  {et~al.}}]{koshida14}
{Koshida}, S., {Minezaki}, T., {Yoshii}, Y., et~al. 2014, \apj, 788, 159

\bibitem[{{Krawczynski} \& {Treister}(2013)}]{krawczynski13}
{Krawczynski}, H., \& {Treister}, E. 2013, Frontiers of Physics, 8, 609

\bibitem[{{Krimm} {et~al.}(2013){Krimm}, {Holland}, {Corbet},
  {et~al.}}]{krimm13}
{Krimm}, H.~A., {Holland}, S.~T., {Corbet}, R.~H.~D., et~al. 2013, \apjs, 209, 14

\bibitem[{{Krist}(1993)}]{krist93}
{Krist}, J. 1993, in ASP Conf. Ser. 52: Astronomical Data Analysis Software and
  Systems II, 536

\bibitem[{{Lira} {et~al.}(2015){Lira}, {Ar{\'e}valo}, {Uttley}, {McHardy}, \&
  {Videla}}]{lira15}
{Lira}, P., {Ar{\'e}valo}, P., {Uttley}, P., {McHardy}, I.~M.~M., \& {Videla},
  L. 2015, \mnras, 454, 368

\bibitem[{{Magorrian} {et~al.}(1998){Magorrian}, {Tremaine}, {Richstone},
  {et~al.}}]{magorrian98}
{Magorrian}, J., {Tremaine}, S., {Richstone}, D., et~al. 1998, \aj, 115, 2285

\bibitem[{{Marinucci} {et~al.}(2014){Marinucci}, {Matt}, {Miniutti},
  {et~al.}}]{marinucci14}
{Marinucci}, A., {Matt}, G., {Miniutti}, G., et~al. 2014, \apj,
  787, 83

\bibitem[{{McConnell} \& {Ma}(2013)}]{mcconnell13}
{McConnell}, N.~J., \& {Ma}, C.-P. 2013, \apj, 764, 184

\bibitem[{{McHardy} {et~al.}(2005){McHardy}, {Gunn}, {Uttley}, \&
  {Goad}}]{mchardy05}
{McHardy}, I.~M., {Gunn}, K.~F., {Uttley}, P., \& {Goad}, M.~R. 2005, \mnras,
  359, 1469

\bibitem[{{McHardy} {et~al.}(2006){McHardy}, {Koerding}, {Knigge}, {Uttley}, \&
  {Fender}}]{mchardy06}
{McHardy}, I.~M., {Koerding}, E., {Knigge}, C., {Uttley}, P., \& {Fender},
  R.~P. 2006, \nat, 444, 730

\bibitem[{{Morris} \& {Ward}(1988)}]{morris88}
{Morris}, S.~L., \& {Ward}, M.~J. 1988, \mnras, 230, 639

\bibitem[{{Netzer}(2015)}]{netzer15}
{Netzer}, H. 2015, \araa, 53, 365

\bibitem[{{Netzer} \& {Laor}(1993)}]{netzer93}
{Netzer}, H., \& {Laor}, A. 1993, \apjl, 404, L51

\bibitem[{{Onken} {et~al.}(2004){Onken}, {Ferrarese}, {Merritt}, {Peterson},
  {Pogge}, {Vestergaard}, \& {Wandel}}]{onken04}
{Onken}, C.~A., {Ferrarese}, L., {Merritt}, D., {Peterson}, B.~M., {Pogge},
  R.~W., {Vestergaard}, M., \& {Wandel}, A. 2004, \apj, 615, 645

\bibitem[{{Onken} {et~al.}(2014){Onken}, {Valluri}, {Brown}, 
  {et~al.}}]{onken14}
{Onken}, C.~A., {Valluri}, M., {Brown}, J.~S., et~al. 2014, \apj, 791, 37

\bibitem[{{Peng} {et~al.}(2002){Peng}, {Ho}, {Impey}, \& {Rix}}]{peng02}
{Peng}, C.~Y., {Ho}, L.~C., {Impey}, C.~D., \& {Rix}, H. 2002, \aj, 124, 266

\bibitem[{{Peng} {et~al.}(2010){Peng}, {Ho}, {Impey}, \& {Rix}}]{peng10}
{Peng}, C.~Y., {Ho}, L.~C., {Impey}, C.~D., \& {Rix}, H.-W. 2010, \aj, 139,
  2097

\bibitem[{{Peterson}(1993)}]{peterson93}
{Peterson}, B.~M. 1993, \pasp, 105, 247

\bibitem[{{Peterson} {et~al.}(2002){Peterson}, {Berlind}, {Bertram},
  {et~al.}}]{peterson02}
{Peterson}, B.~M., {Berlind}, P., {Bertram}, R., et~al. 2002, \apj, 581, 197

\bibitem[{{Peterson} {et~al.}(2004){Peterson}, {Ferrarese}, {Gilbert},
  {et~al.}}]{peterson04}
{Peterson}, B.~M., {Ferrarese}, L., {Gilbert}, K.~M., et~al. 2004, \apj, 613, 682

\bibitem[{{Peterson} {et~al.}(1998{\natexlab{a}}){Peterson}, {Wanders},
  {Bertram}, {Hunley}, {Pogge}, \& {Wagner}}]{peterson98a}
{Peterson}, B.~M., {Wanders}, I., {Bertram}, R., {Hunley}, J.~F., {Pogge},
  R.~W., \& {Wagner}, R.~M. 1998{\natexlab{a}}, \apj, 501, 82

\bibitem[{{Peterson} {et~al.}(1998{\natexlab{b}}){Peterson}, {Wanders},
  {Horne}, {Collier}, {Alexander}, {Kaspi}, \& {Maoz}}]{peterson98b}
{Peterson}, B.~M., {Wanders}, I., {Horne}, K., {Collier}, S., {Alexander}, T.,
  {Kaspi}, S., \& {Maoz}, D. 1998{\natexlab{b}}, \pasp, 110, 660

\bibitem[{{Raimundo} {et~al.}(2013){Raimundo}, {Davies}, {Gandhi}, {Fabian},
  {Canning}, \& {Ivanov}}]{raimundo13}
{Raimundo}, S.~I., {Davies}, R.~I., {Gandhi}, P., {Fabian}, A.~C., {Canning},
  R.~E.~A., \& {Ivanov}, V.~D. 2013, \mnras, 431, 2294

\bibitem[{{Rees}(1984)}]{rees84}
{Rees}, M.~J. 1984, \araa, 22, 471

\bibitem[{{Reynolds} {et~al.}(1997){Reynolds}, {Ward}, {Fabian}, \&
  {Celotti}}]{reynolds97}
{Reynolds}, C.~S., {Ward}, M.~J., {Fabian}, A.~C., \& {Celotti}, A. 1997,
  \mnras, 291, 403

\bibitem[{{Rodr{\'{\i}}guez-Pascual} {et~al.}(1997){Rodr{\'{\i}}guez-Pascual},
    {Alloin}, {Clavel}, {et~al.}}]{rodriguez97}
{Rodr{\'{\i}}guez-Pascual}, P.~M., {Alloin}, D., {Clavel}, J.,
  et~al. 1997, \apjs, 110, 9
    
\bibitem[{{Runnoe} {et~al.}(2012){Runnoe}, {Brotherton}, \& {Shang}}]{runnoe12}
{Runnoe}, J.~C., {Brotherton}, M.~S., \& {Shang}, Z. 2012, \mnras, 422, 478

\bibitem[{{Schlafly} \& {Finkbeiner}(2011)}]{schlafly11}
{Schlafly}, E.~F., \& {Finkbeiner}, D.~P. 2011, \apj, 737, 103

\bibitem[{{Siverd} {et~al.}(2012){Siverd}, {Beatty}, {Pepper}, 
  {et~al.}}]{siverd12}
{Siverd}, R.~J., {Beatty}, T.~G., {Pepper}, J., et~al.  2012, \apj, 761, 123

\bibitem[{{Storey} \& {Zeippen}(2000)}]{storey00}
{Storey}, P.~J., \& {Zeippen}, C.~J. 2000, \mnras, 312, 813

\bibitem[{{Suganuma} {et~al.}(2006){Suganuma}, {Yoshii}, {Kobayashi},
  {Minezaki}, {Enya}, {Tomita}, {Aoki}, {Koshida}, \& {Peterson}}]{suganuma06}
{Suganuma}, M., {Yoshii}, Y., {Kobayashi}, Y., {Minezaki}, T., {Enya}, K.,
  {Tomita}, H., {Aoki}, T., {Koshida}, S., \& {Peterson}, B.~A. 2006, \apj,
  639, 46

\bibitem[{{Tully} {et~al.}(2013){Tully}, {Courtois}, {Dolphin}, 
  {et~al.}}]{tully13}
{Tully}, R.~B., {Courtois}, H.~M., {Dolphin}, A.~E., et~al. 2013, \aj, 146, 86

\bibitem[{{Tully} \& {Fisher}(1977)}]{tully77}
{Tully}, R.~B., \& {Fisher}, J.~R. 1977, \aap, 54, 661

\bibitem[{{Tully} {et~al.}(2009){Tully}, {Rizzi}, {Shaya}, {Courtois},
  {Makarov}, \& {Jacobs}}]{tully09}
{Tully}, R.~B., {Rizzi}, L., {Shaya}, E.~J., {Courtois}, H.~M., {Makarov},
  D.~I., \& {Jacobs}, B.~A. 2009, \aj, 138, 323

\bibitem[{{Urry} \& {Padovani}(1995)}]{urry95}
{Urry}, C.~M., \& {Padovani}, P. 1995, \pasp, 107, 803

\bibitem[{{van Groningen} \& {Wanders}(1992)}]{vangroningen92}
{van Groningen}, E., \& {Wanders}, I. 1992, \pasp, 104, 700

\bibitem[{{Vazdekis} {et~al.}(2010){Vazdekis}, {S{\'a}nchez-Bl{\'a}zquez},
  {Falc{\'o}n-Barroso}, {Cenarro}, {Beasley}, {Cardiel}, {Gorgas}, \&
  {Peletier}}]{vazdekis10}
{Vazdekis}, A., {S{\'a}nchez-Bl{\'a}zquez}, P., {Falc{\'o}n-Barroso}, J.,
  {Cenarro}, A.~J., {Beasley}, M.~A., {Cardiel}, N., {Gorgas}, J., \&
  {Peletier}, R.~F. 2010, \mnras, 404, 1639

\bibitem[{{Walsh} {et~al.}(2009){Walsh}, {Minezaki}, {Bentz}
  {et~al.}}]{walsh09}
{Walsh}, J.~L., {Minezaki}, T., {Bentz}, M.~C., et~al. 2009, \apjs, 185, 156

\bibitem[{{White} \& {Peterson}(1994)}]{white94}
{White}, R.~J., \& {Peterson}, B.~M. 1994, \pasp, 106, 879

\bibitem[{{Whittle}(1992)}]{whittle92}
{Whittle}, M. 1992, \apjs, 79, 49

\bibitem[{{Winkler}(1992)}]{winkler92}
{Winkler}, H. 1992, \mnras, 257, 677

\bibitem[{{Zu} {et~al.}(2011){Zu}, {Kochanek}, \& {Peterson}}]{zu11}
{Zu}, Y., {Kochanek}, C.~S., \& {Peterson}, B.~M. 2011, \apj, 735, 80

\end{thebibliography}

\end{document}